\documentclass[11pt]{article}

\usepackage{amsmath}
\usepackage{amssymb}
\usepackage{graphicx}
\usepackage{dcolumn}
\usepackage{bm}
\usepackage[margin=2cm]{geometry}
\usepackage{hyperref}
\usepackage{pgf,pgfarrows,pgfnodes,pgfautomata,pgfheaps,pgfshade}
\usepackage{graphicx,amsfonts}
\usepackage{amsmath,amssymb,setspace}
\usepackage{mathrsfs}
\usepackage{colortbl}
\usepackage{tikz}
\usepackage{framed}
\usepackage{subfig}
\usepackage{float}
\usepackage{enumerate}
\usepackage{epstopdf}
\usepackage{graphicx}
\usepackage{geometry}
\usepackage{lipsum}  
\usepackage{color}

\numberwithin{equation}{section}
\numberwithin{subsection}{section}
\newtheorem{thm}{Theorem}[section]

\newtheorem{proposition}[thm]{Proposition}

\newenvironment{proof}{{\bf Proof.}}{\hfill$\square$\vskip.5cm}

\newcommand{\Addresses}

\title{Go With the Flow, on Jupiter and Snow.
\\
Coherence From {\color{black} Model-Free} Video Data without Trajectories.}

\author{
  {Abd AlRahman AlMomani$^{1}$ and Erik M. Bollt$^{2}$} \\
  {$^{1,2}$ Department of Mathematics and Computer Science, Clarkson University,}\\{ Potsdam, New York 13699, USA.} \\
  {$^{1}$ almomaa@clarkson.edu,  $^{2}$ bolltem@clarkson.edu}
  }

\date{}

\begin{document}

\maketitle


\begin{abstract}



Viewing a data set such as the clouds of Jupiter, coherence is readily apparent to human observers, especially the Great Red Spot, but also other great storms and persistent structures.  There are now many different definitions and perspectives mathematically describing coherent structures, but we will take an image processing perspective here.
We describe an image processing perspective inference of coherent sets from a fluidic system directly from image data, without attempting to first model underlying flow fields, related to a concept in image processing called motion tracking.  In contrast to standard spectral methods for image processing which are generally related to a symmetric affinity matrix, leading to standard spectral graph theory, we need a not symmetric affinity which arises naturally from the underlying arrow of time.
We develop an anisotropic, directed diffusion operator corresponding to flow on a directed graph, from a directed affinity matrix developed with coherence in mind, and corresponding spectral graph theory from the graph Laplacian. {\color{black}Our methodology is not offered as more accurate than other traditional methods of finding coherent sets, but rather our approach works with alternative kinds of data sets, in the absence of vector field}.  Our examples will include partitioning the weather and cloud structures of Jupiter, and a local to Potsdam, N.Y. lake-effect snow event on Earth, as well as the benchmark test double-gyre system.

\end{abstract}

%


\section{Introduction}

There has been a significant emphasis in recent dynamical systems literature to define, and find ``coherent structures" \cite{shadden, Bovens, Fitelson, Warfield, Merricks, Shogenji, Froyland, Froyland2,Froyland4}.  It could be said that these methods could be divided into those that follow interiors of sets by transfer operators, or those that define a property of boundaries of such sets and follow boundary curves \cite{bolltCurv,bolltCurv0,HallerA,HallerB}.  Some methods have been developed and put forward without specifically defining the coherence principle that the method is designed to extract.  In any case, perhaps most would agree that coherence should be defined in some manner to describe sets (of particles) that ``hold together" for some time, or densities of ensembles of particles \cite{Yorke, Neumann, Mori}, or measurements thereof \cite{appliedkoopmanism, LinearizationKoopman, william, shervin}.
{\color{black}However, we will present a perspective of coherence regarding a pattern that persists in time, whether or not the underlying advecting particles hold together.  This is a not necessarily Lagrangian perspective that makes sense in terms of asking what is measured, and which we highlight by our examples.
 }

In essentially all of the studies that have appeared in recent literature, no matter what the method, approach or perspective, one starts with a dynamical system.  From there follows the quantity to be analyzed.  In other words, an underlying flow is assumed in the sense that generally a differential equation is required to proceed, whether explicitly or implicitly through observations of an experiment.  For this, we will write, 

\begin{equation}\label{model}
\dot{{\mathbf x}}={\mathbf F} ({\mathbf x},t),
\end{equation}

\noindent for a vector field, ${\mathbf F}:M\times {\mathbb R} \rightarrow M$, (typically $M\subset \mathbb{R}^2$ or perhaps $\mathbb{R}^3$), but this may be developed from a stream function from an underlying partial differential equation for example.  In any case, then a flow mapping, ${\mathbf x}(t) =\Phi({\mathbf x}_0,t_0,t)$ is inferred, even if this means numerical integration of the differential equation.  

In recent work, aspects of advection and diffusion have been both involved in developing a better understanding of coherence \cite{Froyland,Froyland2, Junge}, including for models of stochastic processes.
We summarize that universally, previous work  either begins with a model of the dynamical system, or at least attempted to empirically develop a model perhaps by optical flow, including our own \cite{bolltOptic1, bolltOptic2, bolltOptic3} or similarly by other means \cite{haller}, and recently by Koopman operator methods but requiring a vector field \cite{Poje}.

In contrast to all the mathematical formalism and machinery behind current studies of coherence, it can be said that people ``recognize" coherent sets when they see them; consider that the Great Red Spot of Jupiter is clear to any and all that have seen it, as perhaps the most famous coherent set in the solar system.  With this motivation, we will develop here an observer based perspective of coherence.

If we do not have a model, as the dynamical system is known only by remote sensing observations, then in practice  the flow mapping, $\Phi({\mathbf x}_0,t_0,t)$ is at best inferred, but generally not available, and often likewise nonlinear systems require numerical integration to infer the flow at sampled points.  Here we will approach questions of coherence in the setting that we have only remote observations, but no model.
Developing a model of the flow either directly, $\Phi({\mathbf x}_0,t_0,t)$, or as a model of the vector field (say by optical flow), may not always be practical or the best way to proceed.\\

\begin{figure}[H]
  \centering
\includegraphics[width=.45\textwidth]{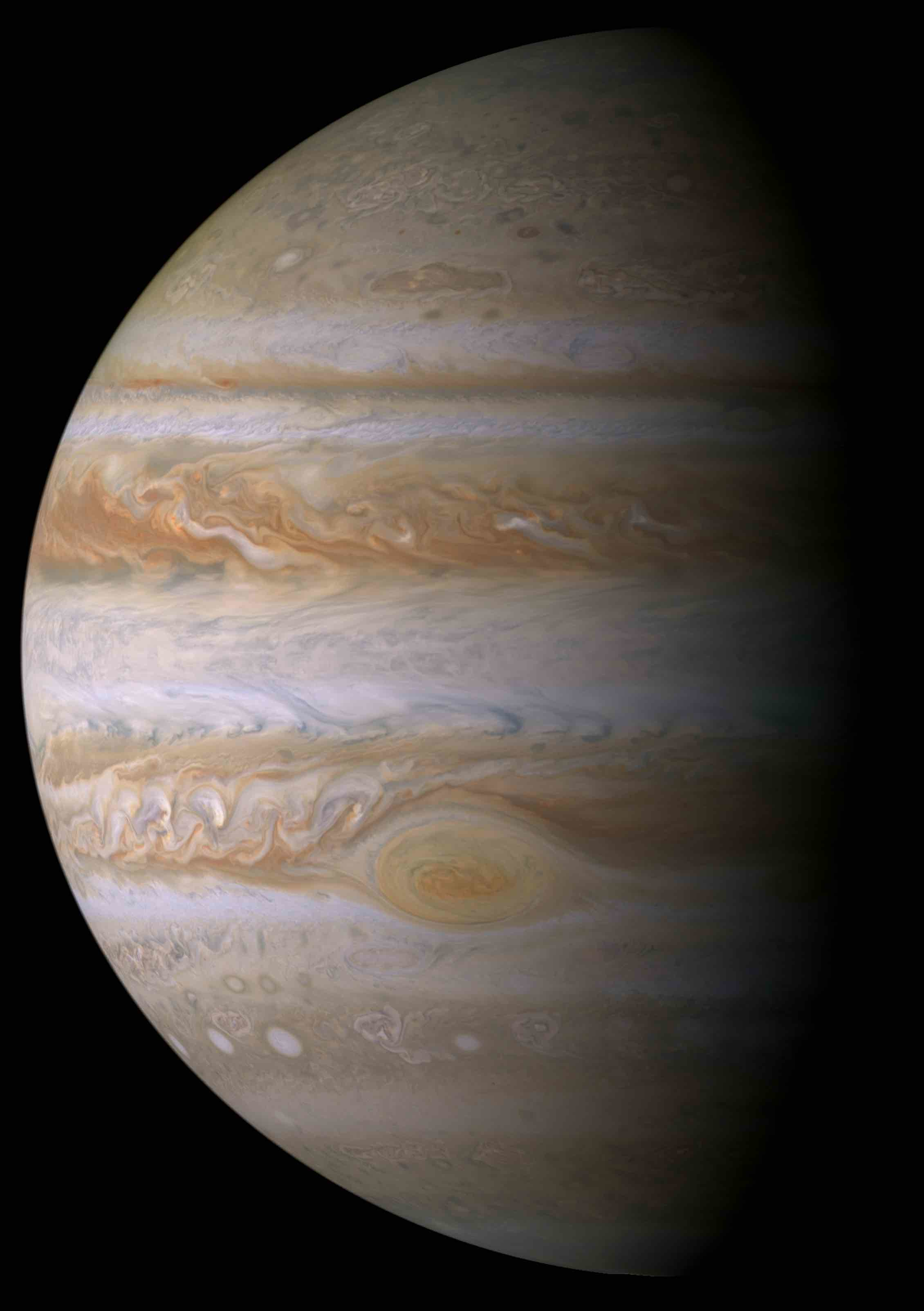}              
  \caption{Jupiter Portrait as viewed from the spaceship Cassini. ``This true color mosaic of Jupiter was constructed from images taken by the narrow-angle camera onboard NASA's Cassini spacecraft on December 29, 2000, during its closest approach to the giant planet at a distance of approximately 10 million kilometers (6.2 million miles)." \cite{nasacite}. 
  }
  \label{fig1}
\end{figure}
 
 \begin{figure}[ht]
  \centering 
\includegraphics[width=.6\textwidth]{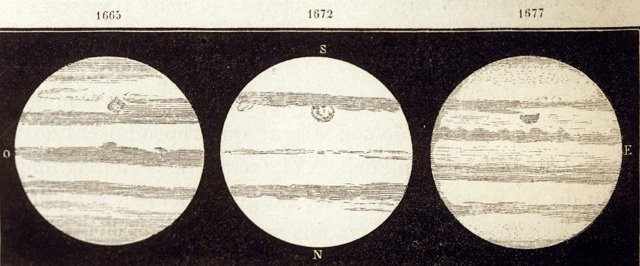}              
\includegraphics[width=.6\textwidth]{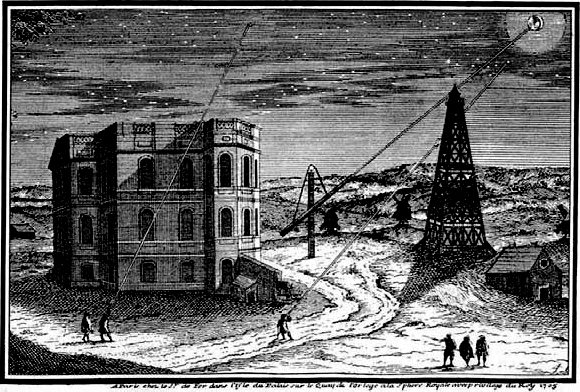}              
  \caption{Jupiter as sketched by Giovanni Domenico Cassini (Top) in his own hand from 1665-1677, from the Memoires de l'Acad\'{e}mie Royale des Sciences de Paris \cite{discovery}.  Note that North is drawn, and so labeled, on the bottom.  We see that Cassini was seeing and sketching similar scenes over the several years, including apparently the large storm.  (Bottom) A sketch of the observatory in Paris.
  }
  \label{fig2}
\end{figure}

{\color{black} The Great Red Spot (GRS), for example,  was observed and identified as persistent since 350 years ago, initially by human eye observations without a great deal of formalism associated with our modern descriptions and algorithms of coherent sets.}
Note that this historical observation was made absent transfer operators methodology \cite{Yorke, Mori}, and likewise absent Koopman operators formalism \cite{appliedkoopmanism, william}, and also even optic flow methods were not used \cite{bolltOptic3}.  
See Fig.~\ref{fig1}, as seen in the year 2000 from the Cassini space probe, a joint NASA, European Space Agency (ESA), and Italian space agency Agenzia Spaziale Italiana (ASI) mission \cite{cassini2000}. 
The solar system's largest and most persistent planetary hurricane storm, the vortex structure called the GRS is clearly visible in this image. 
There are also belts and zones as persistent latitudinal structures, as well as many other smaller storms, (but still massive by Earth standards).  There are also other embedded objects, that are clearly present and notable by the naked eye, without ever needing a digital computational engine {\color{black}to identify. It is as }clear today to the casual observer of these modern images, as it was to the Renaissance era astronomer Giovanni Domenico Cassini himself that there are coherent structures on Jupiter \cite{discovery}.  See Fig.~\ref{fig2} where Cassini's sketches show some of the same structures as viewed across several years from 1665-1677, were clear enough that he was able to see them despite what were low-quality telescopes by any modern standard, and many of these structures persist today, hundreds of years later.  It is important to distinguish between the concept of a feature that we may notice in a single image as compared to a feature that persists over several successive images, over time.  Persistence over time is more akin to what is meant by coherence; we will contrast image segmentation concepts versus motion segmentation concepts.  In Sec. \ref{directedspectralpartition},  this contrast leads us to a directed affinity.

{\color{black}  \textbf{It is our goal here} to develop an observer based perspective of coherence, and to define the manner in which we {\it know} a coherent structure when we see it, by appropriate mathematical formalism.}  {\color{black}Note that our methodology is not offered as computationally ``better" or more accurate than other traditional methods of finding coherent sets. Rather our approach works with alternative kinds of data sets.  Specifically, we do not require vector fields, or trajectories of particles describing the underlying dynamics. Instead, we work model free directly with observational frames (video).  So while our results do appear similar, since the underlying assumptions are different, they may not be identical.}

Furthermore, and also central to this work, since an observer can gain impression of persistence of certain structures, then our perspective should be developed to be directly comparable to imagery, without needing to go through steps of modeling the imagery by computed vector fields, then integrating the vector fields numerically to develop a flow map, before only then involving the methods of geometric dynamical systems; note that we admit this is counter even to our own previous efforts \cite{bolltOptic1, bolltOptic3} which have followed this exact prescription for data driven remote sensing starting with specializing optical flow methods and comparably even for the study of the atmospheres of Jupiter by \cite{HallerLAST}.  The principle we choose in this current work is that we could start and end with the images themselves, as representing pointwise measurements in time.

{\color{black}In this paper, we structure the presentation as follows.  In Sec.~\ref{measealongorbits} we relate to the concept that observations in a color image are spatial measurements, which when evolved in time, relate to measurements advected along orbits, noting an observer-centric perspective.  We review in Sec~\ref{ImSeg}, the especially popular methods of k-means clustering and also spectral clustering, as related to image segmentation.
In Sec.~\ref{directedspectralpartition} we formalize the idea of motion segmentation as a partition in space, along time, and we relate this to coherency.  We contrast ideas from the image processing community about image segmentation which is inherently a symmetric concept in most algorithmic approaches one might take, as compared to motion segmentation that leads necessarily to a not symmetric description due to the arrow of time inherent in the concept. Then, we develop a directed affinity which naturally incorporates the asymmetry as directed by time in a manner that describes coherency as a spatial and time oriented version of ``particles hold together", with details relating to the graph Laplacian of a \textit{directed graph} underlying the directed graph version of spectral graph theory.   
In Sec.~\ref{numericalresults}, we address experimental and numerical results by showing motion segmentation by directed spectral segmentation method that naturally finds convincing coherent sets, in data from Jupiter, from a data set from a local lake effect  storm near our own university, and from the highly popular double-gyre system often used for benchmarking coherent set analysis.
In Sec.~\ref{cutSymmetric}-\ref{directedspectralGraph} we include background material regarding the nCut problem, for clustering, and leading to spectral clustering for the \textit{directed graph problem} with the corresponding special case of a \textit{weighted directed graph Laplacian}, as used in the \textit{directed spectral segmentation}.}

\section{Measure Along Orbits}\label{measealongorbits}
{\color{black} Suppose the dynamical system, Eq.~\ref{model}, over the phase space $M$ may not be directly known to us, and we have an observation that is Lebesgue measurable,}

\begin{equation}
{\mathbf c}:M  \rightarrow {\mathbb R}^d,
\end{equation}

\noindent where $d$ is the number of scalar measurements made.  In the case of the image shown in Fig.~\ref{fig1}, $d=3$, and ${\mathbf c}$ samples three color intensities from a standard color scale (such as RGB as shown) at each point ${\mathbf z}$ in the field of view; that is at a given time, $c_j({\mathbf z}):M\rightarrow {\mathbb R}$, $j=1,2,3$, measures any one of the color plane values at ${\mathbf z}$.  

For intrinsic and not directly measured quantities, {\color{black} $h_i:{\mathbb R} \times M\rightarrow {\mathbb R}$, $h(t,z)$ describes  pressures, temperatures, or gas chemical concentrations, but we abuse notation to write $h_{i}(z) \equiv h(t_{i},z)$.}
{\color{black}In the case of Jupiter, for example,  the colors and intensities at each point represent chemical concentrations of various chemicals in the clouds, densities, depth of the cloud layers, and other properties as inferred by reflectance \cite{JupiterColors}. 
Then measured quantities are a collection of functions $\{h_i\}_{i=1}^K$, hidden to us but combined into the function, ${\mathbf c}({\mathbf z})={\mathbf q}(h_1({\mathbf z}),..,h_K({\mathbf z})),$ by some unknown to us function ${\mathbf q}$ related to the underlying physics.}

What allows us to describe  patterns in images as  coherent, is that they persist in some form across many frames of the ``movie", meaning as the system is observed through multiple times, and gradually evolving, what is seen is close enough to the original that we recognize it; in \cite{bolltCurv0,bolltCurv} we suggested the concept of shape coherence as sets that almost maintain shape over time.  

{\color{black}In \cite{Froyland6,Froyland4}, a concept} of coherent pairs was developed that roughly states that a coherent pair of sets $A$ and $B$ should be such that $\Phi(A)\approx B$, but also $\Phi^{-1}(B)\approx A$ with some notion of diffusion or randomness to reward those set pairs when the boundary does not grow too large.
{\color{black}We have simplified the notation of Eq.~\ref{model}, as {\color{black} ${\mathbf z}=\Phi(z_0)=\Phi({\mathbf z}_0,t_0,t)$ } suppressing the notation  $t_0,$ $t$, for sake of simplicity. }

The idea of studying the boundary of sets then also relates to the concepts of recent formulations of geodesic Lagrangian Coherent Structures (LCS) and transport barriers in terms of studying  strain and also length minimizing curves \cite{HallerA,HallerB}.  In some sense, both aspects of stretching and folding associated with curvature may have a role \cite{bolltCurv1}.

{\color{black}It is interesting to relate the concept of observation, as described here, to the notion of measuring along orbits related to the Koopman operator \cite{appliedkoopmanism}.  Considering ${\mathbf z}=\Phi({\mathbf z}_0)$ as the ``down stream" image of an initial condition ${\mathbf z}_0$, then to measure (the colors) down stream from ${\mathbf z}_0$ is a concept defined by the Koopman operator formalism \cite{appliedkoopmanism,LinearizationKoopman} which we recall\cite{appliedkoopmanism, bolltbook}, ${\cal K}:{\cal F}\rightarrow {\cal F} \nonumber,
{\cal K}[h]({\mathbf z})=h\circ \Phi({\mathbf z})$,
 where ${\cal F}$ may be taken as, $L^\infty(M)$.  } 
Recall that, ``...the Koopman operator maps functions of state space to functions of state space and not states to states" \cite{william}.     
Several measurable functions such as 
${\cal G}=(g_1,..,g_K)$, 
has been called, a ``vector valued observable," \cite{william}.  A Koopman operator applied to each is inherited by the vectorized version of the Koopman operator, ${\cal K}_t[{\mathbf c}]=({\mathbf K}_t[c_1],..,{\mathbf K}_t[c_k])$.

{\color{black}Measuring color at a set of points $A$, may be written as ${\mathbf c}(A)$, as a remote measurement of the scene, related to gas chemical concentrations, pressures and so forth.  Measurement downstream would be, {\color{black} ${\cal K}[{\mathbf c}](A)$ } as the push forward of $A$.}  {\color{black}Since the adjoint ${\cal K}^*$ has properties of a pull back operator, (and associate with the Frobenius-Perron transfer operator, for functions in $L^1(M)$ \cite{bolltbook}), then to have measurements on the push forward match those on the pull back, is to demand approximately, ${\cal K}^*{\cal K}[{\mathbf c}](A)\approx A$.  Likewise, stated in reverse, there should be an approximate ``coherence" matching ${\cal K}{\cal K}^*[{\mathbf c}](B)\approx B$.}  Note then that, ${\cal K}^*$ may be defined in terms of the pull back ${\cal K}^*[\rho]({\mathbf z})=\rho\circ \Phi^{-1}({\mathbf z})$, at least when $\Phi^{-1}$ exists as it will if it is a flow, and is measuring preserving, but alternatively, the Frobenius-Perron operator is, ${\cal K}^*[\rho]({\mathbf z})=\int_M \delta({\mathbf z}-\Phi({\mathbf y})) \rho({\mathbf y} d \mathbf y$, and the Koopman operator can be written, ${\cal K}[h]({\mathbf z})=\int_M \delta({\mathbf y}-\Phi({\mathbf z})) h({\mathbf z} d \mathbf y$, where for sake of brevity, we are suppressing statement of the space of functions where this is appropriate, and the corresponding discussion of bilinear forms relating the operator and its adjoint \cite{appliedkoopmanism, bolltbook}.

In \cite{Froyland6} a spectral method was developed {\color{black}associated} with these eigenfunction type statements for the operator, ${\cal K}^*{\cal K}$, and this description is expanded upon in \cite{Koltai}.  An average of both forward and backward time coherent pairings was {\color{black}offered} in \cite{Froyland7}, including a statement that these concepts are associated with keeping small boundaries.  {\color{black}These have proven to be very effective and powerful approaches}, however, they require Lagrangian trajectory data.  

Even recent clustering methods such as the k-means approach in \cite{Froyland8}, or the spectral approach in \cite{HallerC} require Lagrangian trajectory data (stated roughly as measurements following along with orbits).  There has been related work in spatiotemporal feature extraction and forecasting from the Koopman perspective \cite{Giannakis1, Giannakis2}, but also adaptations of the Koopman operator for image texture analysis\cite{Surana} and also for video segmentation \cite{Chan}. 
  It contrast, it can be said that remotely sensed ``movie" data is inherently Eulerian (stated roughly as measurements associated with fixed positions in space).

With this background, we now proceed to contrast image segmentation methods toward developing a spectral motion segmentation method.
Notice that when only movie data is available, then we specifically lack the Lagrangian trajectory data to explicitly carry forward any of the several operator methods or boundary methods or  LCS methods from the literature. 

So in this case, we proceed to build a proxy operator, that rewards concepts of like measurements and close distance, and in many ways, this proxy operator serves the role of a transfer operator estimator, perhaps likely a Bayesian estimator, which we plan to pursue as a question in future work.  Only the DMD method (Dynamic Mode Decomposition) \cite{appliedkoopmanism} \cite{LinearizationKoopman}-\cite{rowley} can also directly handle movie data, but is also somewhat different in approach to how the operator is estimated by the least squares approach. As for now, notice that stated as an anzatz, we are emphasizing continuity in space and continuity in time measurements of the underlying but unknown flow.

\section{Image Segmentation and Symmetric Affinity, versus Motion Segmentation and Not Symmetric Affinity} \label{ImSeg}

\begin{figure}[htbp]
  \centering
\includegraphics[width=.6\textwidth]{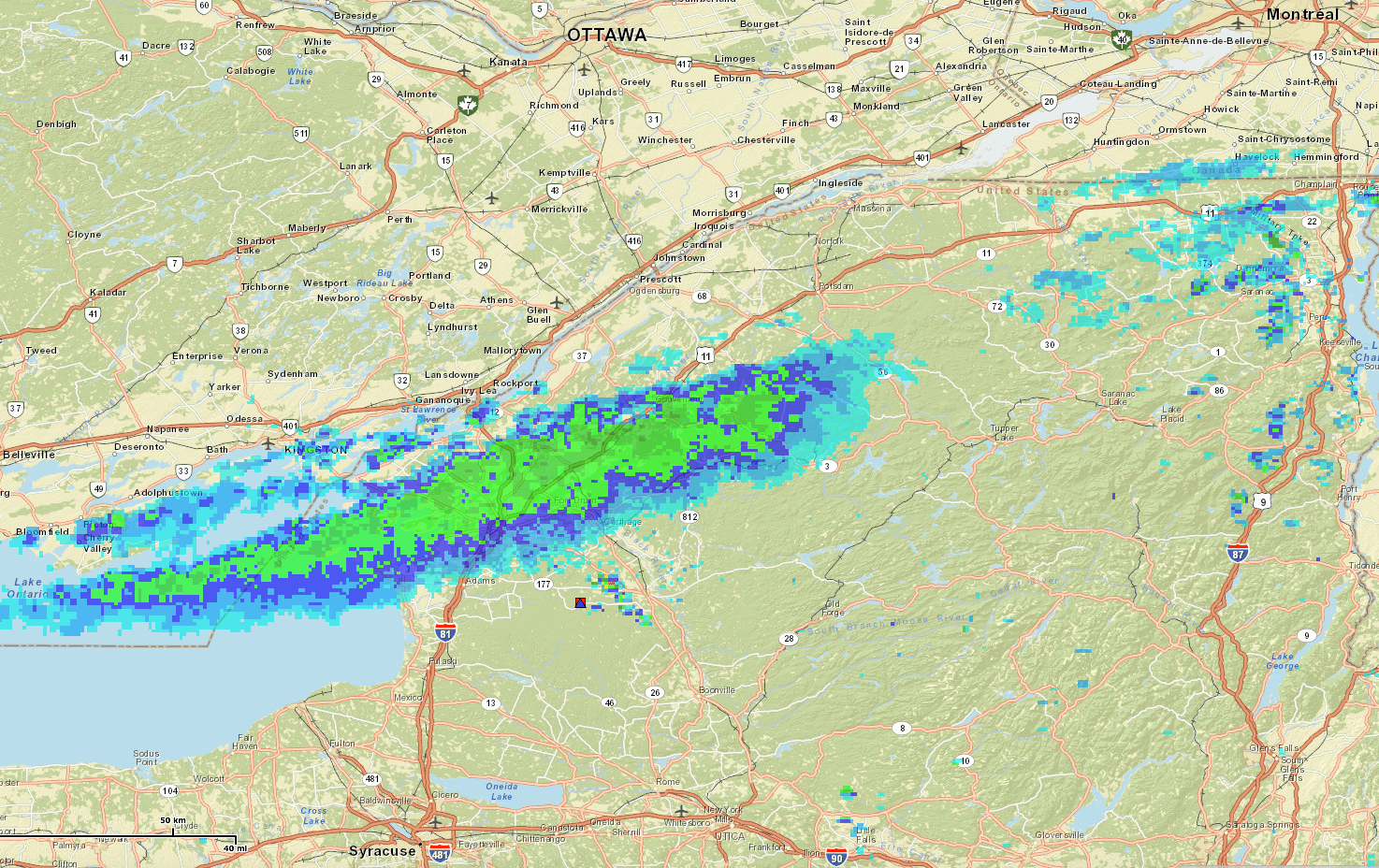}              
  \caption{{\color{black} Lake effect snow on NEXRAD level III National Reflectivity Mosaic and Data radar imagery, near Watertown, NY indicating precipitation of a period of intense lake effect snow during a 4-day period from Nov 18-21, 2014, shown here on 2014/11/18 at 12EST.  The northern side of this localized storm skirts near Clarkson University, Potsdam, NY, location of the writing of this article.  Syracuse, NY, Ottawa, ON, and Montreal, ON are shown for map perspective. The striking feature of a lake effect snow event from the view of NEXRAD is the energetic snow that seems to stream off the lake but does not move downstream with other tracers and evidence that moves more so with the underlying advection of particles in the flow.}
  }
  \label{fig3}
\end{figure}

``Following along measured observations," for clustering is not necessarily the same as following along orbits for coherence, but they are easily confused, even if these ideas sometimes may coincide.  And rightly, they both might be called coherence depending on the setting.    We have quoted the phrase ``following along measured observations," because this is roughly describing a cluster of like measurements that tracks in time, but may not specifically be exactly attached to the underlying advective flow. 


In the image processing community, the general problem of clustering ``like objects" between frames of a movie is called {\it motion segmentation}, also known as {\it image motion segmentation}, and in turn, the outcome of these have been used in the image processing community to infer {\it motion tracking}.

Motion tracking of objects, or tokens, is not necessarily the same problem as the inference of the underlying flow  ({\it tracking~\cite{motionSeg} would traditionally be applied in image processing of say a movie of people moving to reveal the underlying motion of individuals, or groups of individuals, as a ``token".}) which is easily seen when considering the weather event illustrated in Fig.~\ref{fig3}.  We will cast this work as one of motion tracking and then contrast to the Lagrangian coherent structures problem \cite{haller3}.  We argue that only the former is truly accessible by remote sensing.    

First, we review the static time problem of {\it image segmentation,} generally as clustering problems, and in the language of our image data from remote sensing.
Consider  clustering within a single scene, meaning a single frame of a movie.
Suppose a grid of positions where color (or some other collection of pointwise measured quantities) is sampled, at each of $\{{\mathbf z}_{i}\}_{i=1}^{M}$ for $M$ (usually uniformly spaced grid of) pixels over $\{{\mathbf z}_i\}_{i=1}^{M}\subset {\cal M}\subset {\mathbb R}^2$.  So ${\cal M}$ is the framed image.
At each of these, observe ${\mathbf c}({\mathbf z}):{\cal M}\rightarrow {\mathbb R}^d$, (generally say $d=3$ colors at each position) to form an observation matrix, 

\begin{equation}
{\color{black} X_{i,:} }={\mathbf c}({\mathbf z}_{i}).
\end{equation}

Since ${\mathbf c}$ is a vector valued measured observation with $d$ observations (colors), then $X$ is $M \times d$.  
For many frames sampled across time, {\color{black} we will write, $X_{i,r}[q]$,where  $1\leq r\leq d$ ``colors", and for each time $t_q$, $0\leq q\leq N-1$}. 

Our goal is to partition the space based on a notion of coherency, across time and space.  By spatial partition the space of sampled data, we mean, given data $\{{\mathbf z}_{i}\}_{i=1}^{M}$, there  is an assignment into labels, {\color{black} ${\cal S}=\{{\cal{S}}_l\}_{l=1}^k$}  that serves as a function from the pixel positions to (colored) labels.  How this assignment should be done appropriately is a matter we now discuss and we describe how it should relate to how the measured ${\mathbf{c}}$ values vary across time. {\color{black} In other words, since each discretely indexed position ${\mathbf z}_{i,j}$ gets assigned one of $k$-levels, a membership in each ${\cal{S}}_l$ can be represented by a unique color, thus recoloring the space by those $k$-levels serves to partition the space, also known as a segmentation.}  

Usually a clustering is useful if it associates ``like" (in some suitable sense) ${\mathbf c}$-measured values of the data.
Perhaps the two most commonly useful image segmentation methods are called $k$-means\cite{k-means} and spectral segmentation\cite{NgJordanWeiss} respectively. 

Image segmentation may be formulated as a spectral graph partitioning problem \cite{NgJordanWeiss}, which we review in Sec.~\ref{spectralcoloralone}.
However, these methods need a major adjustment when applied to image sequences (movies) for motion segmentation, despite that traditionally they have been applied to movies with some degree of success \cite{Shi-Malik}.  The key difference is what underlies a notion of coherent observations, remembering that the arrow of time has directionality. We require {\it affinity matrices that are not symmetric, }and when considered as graphs, they are {\it directed graphs.}  Therefore much of the theoretical underpinning of the standard spectral partitioning needs some adjustment, since it relies on symmetric matrices and undirected graphs.  We will need a graph Laplacian for weighted directed graphs.

%

\subsection{On k-means Image Segmentation by Color Alone}
\label{k-meansalone}
A simple and common form of clustering that one might choose would be a k-means clustering of an image scene \cite{k-means} based just on the pointwise measurements alone (say colors for example) as a solution to the partitioning problem, to find a partition ${\cal S}$ such that
{\color{black}
\begin{equation}\label{k-meansApp}
{\cal S}=\underset{S}{\operatorname{argmin}} \sum_{l=1}^k \sum_{j \in {\cal S}_l} \| X_{j,:}-\mu_l\|,
\end{equation}}
\noindent where $\|\cdot\|$ is the Euclidean norm of the color values, and {\color{black}$\mu_l$} are means in each color channel {\color{black}  ${\cal{S}}_l$}.
We see the k-means method is a solution of a partitioning problem.
An image such as that of the colors of Jupiter is shown in Fig.~\ref{fig10} (b and c) for an example of a static time segmentation of a Jupiter image with $d=3$ colors ${\mathbf c}({\mathbf z})$, measured pointwise where {\color{black} $\{{\mathbf z_i}\}_{i=1}^{M}$ } are the pixel positions on the image.  
The k-means problem solution has a direct method of updating the cost function {\color{black} Eq.~\ref{k-meansApp} } as membership of indexed values in each partition element is adjusted, {\color{black}optimizing relative to shifting the group mean}, as reviewed in many standard texts \cite{k-meansasspectral, k-means,Arya}.



\subsection{On Spectral Segmentation by Color Alone}
\label{spectralcoloralone}

There have been several complementary views of clustering by spectral methods, by
graph cuts \cite{Shi-Malik, Kannan-Vapela}, as random walkers \cite{Meila-Shi}, and comparably as a diffusion process as described by diffusion map \cite{diffusionmap} and comparably as an eigensystem.  Many of these come back to some version of a max-flow min-cut algorithm that we will review in Sec.~\ref{cutSymmetric} \cite{max-flow}, and in turn as related to the conductance also called Cheeger-constant as a measure of ``bottleneckyness" of the underlying graph.  In this section, we review the computations for the simpler case of weighted {\it undirected} graphs, appropriate for image segmentation, but in the subsequent section, we will relate our motion segmentation problem to the graph problem of weighted {\it directed} graphs to account for the directed aspect of the arrow of time.

Proceeding computationally, image segmentation may be formulated as a graph partitioning problem, and as such, doing so with color alone means formulating the data set; assign data set \cite{NgJordanWeiss},
\begin{equation} \label{eq:X}
X=[X_{1,:}^T|X_{2,:}^T|...|X_{M,:}^T].
\end{equation}
So, for color alone, $X$ is $d\times N$.  Columns of $X$ are the color channels at each pixel position ${\mathbf z}_i$, and we write $X_i=X_{i,:}^T$.  If distance is based on color alone, and so as in {\color{black} Eq.~\ref{k-meansApp} }, we write a
pairwise distance function. Let
\begin{equation}\label{justimage}
D_{i,j}=\|X_i-X_j\|=\sqrt{{\color{black}\sum_{k=1}^d} (X_{i,k}-X_{j,k})^2},
\end{equation}
\noindent describe a matrix of distance function values across the sample of points, for distance function, $d({\mathbf z}_i,{\mathbf z}_j)$, and $d:M\times M \rightarrow {\mathbb R}^+$.
%
Next as done in many general spectral clustering methods \cite{NgJordanWeiss, diffusionmap} and as specialized to image segmentation \cite{Shi-Malik,NgJordanWeiss},  a pairwise symmetric affinity matrix  may be defined,
\begin{equation}\label{symmetricW}
W_{i.j}=e^{-D_{i,j}^2/2\sigma^2}.
\end{equation}
The value of $\sigma>0$ may be chosen as a resolution parameter. It is convenient to emphasize the  ``practical" sparsity, by reassigning $W_{i,j}=0$ if $W_{i,j}<\epsilon$ for  a small threshold, $\epsilon>0$.
This can be interpreted as generating a weighted graph, $G=(V,E)$, where vertices {\color{black} $V=\{1,2,...,M\}$ } have edges between them whenever $W_{i,j}>0$ and with weights accordingly.

A degree matrix, corresponding to the weighted symmetric directed graph is,
\begin{equation}\label{degreematrix}
{\cal D}(i,i)=\sum_j W_{i,j}, \mbox{ }{\cal D}_{i,j}=0, i\neq j.
\end{equation}
Shi and Malik \cite{Shi-Malik} realized noted that the  max-cut, (see Appendix \ref{cutSymmetric} ), is equivalent to,
\begin{equation}
\min_x ncut(x)=\min_y \frac{y^T ({\cal D} -W)y}{y^T{\color{black}{\cal D}}y},
\end{equation}
as can be proved through the Courant-Fischer theorem \cite{bolltbook}, and  \cite{Shi-Malik} for the image segmentation setting.
This then brings us to the generalized eigenvalue eigenvector problem, 
\begin{equation}\label{eigprob}
({\cal D} -W)y=\lambda {\color{black}{\cal D}} y,
\end{equation}
where the second smallest eigenvalue and corresponding eigenvector solve the optimization problem.  This could be written in terms of a symmetric normalized graph Laplacian, $L$, by noting that Eq.~\ref{eigprob} transforms into,
\begin{equation}
{\cal D}^{-1/2}({\cal D} -W){\cal D}^{-1/2}x=\lambda x,
\end{equation}
 or,
 \begin{equation}
 Lx=\lambda x,
 \end{equation}
  if,
 \begin{equation}
 L= {\cal D}^{-1/2}({\cal D} -W){\cal D}^{-1/2},
 \end{equation}
  by substitution,
  \begin{equation}\label{eigprobfinish}
  y={\cal D}^{-1/2}x.
  \end{equation}
  
The affinity matrix eigenvalue problem has an interpretation as a stochastic matrix eigenvalue problem, by  \cite{Meila1,Meila-Shi},
\begin{equation}
P={\cal D}^{-1}W.
\end{equation}
Meila and Shi \cite{Meila-Shi} noted that the affinity matrix $W$ relates to random walks through a graph according to this stochastic matrix  $P$, and this relates closely to a diffusive process underlying the diffusion map method \cite{diffusionmap,diffusionmap2}.
This random walker interpretation connection between eigenvalues of $P$ and $W$ is reviewed further in Sec.~\ref{randomwalkapp}

Now the smallest eigenvalue of Eq.~\ref{eigprob} corresponds to the greedy partition (when partitioning a graph into two sets A and B, one element of the two partitions is empty) so the second smallest eigenvalue corresponds to the Cheeger-balanced partition, the best bi-partition.  Then one could proceed by recursively bi-partitioning \cite{TianErik}.  We follow the concept of \cite{NgJordanWeiss} which is to choose the $k$ smallest eigenvalues {\it after the zero eigenvalue} and corresponding eigenvectors and then to cluster these by use of k-means from there.  This is what we see in Fig.~\ref{fig10}(b).


\section{Motion Segmentation, and Directed Affinity, Following Along Measured Observations}
\label{directedspectralpartition}

Now we develop a directed affinity matrix ${\cal W}$ (note the change of font to distinguish from the symmetric counterparts $W$ in Eq.~\ref{symmetricW}). Replace $X$ in Eq.~\ref{eq:X} with, 
\begin{equation}
X(t)=[X_{1,:}(t)^T|X_{2,:}(t)^T|...|X_{M,:}(t)^T],
\end{equation}
where $X_{i,:}(t)$ denotes the column vector of $d$ colors at ${\mathbf z}_i$, pixel location $i$, at time $t$ in the movie sequence.   Generally the colors at pixel $i$ will be changing over time.  
Then let,
\begin{eqnarray}
D_{1}(i,j,a,\tau)&=&
\sum_{l=0}^{\tau-1}\|X_i(t+l a)-X_j(t+(l+1)a)\| \nonumber \\
&=& \sum_{l=0}^{\tau-1}\sqrt{{\color{black}\sum_{k=1}^d} (X_{i,k}(t+(l)a) -X_{j,k}(t+(l+1)a))^2}.
\end{eqnarray}
This compares the scene at pixel position $i$, through $\tau$-time instances, $ l=0,a, 2a,...,  (\tau-1) a$, to the  scene at pixel $j$ through $\tau$-time instances one step in the future,  $ l=1,a, 2a,...,  \tau a$.
Note that the norm, the inner sum, is the same as the color measuring norm in Eq.~\ref{justimage}.

Now we measure the spatial distance between the pixels, as they appear naturally in the scenes represented by the figures.  Let,
\begin{equation}
D_2(i,j)^2=\|{\mathbf z}_i-{\mathbf z}_j\|^2=(z_{1,i}-z_{1,j})^2+ (z_{2,i}-z_{2,j})^2,
\end{equation}
where ${\mathbf z}_i=(z_{1,i},z_{2,i})$ denotes the spatial coordinates of pixel number $i$.  This is the standard spatial Euclidean norm.

Adding these two norms defines a spatial and time delayed color distance function,
\begin{equation}\label{affinityd}
D(i,j,a,\tau)^2=D_{1}(i,j,a,\tau)^2+\alpha D_2(i,j)^2.
\end{equation}
Finally an affinity matrix follows,
{\color{black}
\begin{equation}\label{affinity2}
{\cal W}_{i,j}=e^{-D(i,j,a,\tau)^2/2\sigma^2}.
\end{equation}
}
Notice we have suppressed including all the parameters in writing ${\cal W}_{i,j}$, and that besides time parameters $a$ and $\tau$ that serve as sampling and history parameters, together the parameters $\alpha$ and $\sigma$ serve to balance spatial scale and resolution of color histories, and comparable to the role of $\sigma$ in Eq.~\ref{symmetricW}. 

Contrasting $W$ in Eq.~\ref{symmetricW} to ${\cal W}$ in Eq.~\ref{affinity2} {\color{black}we see the difference} of symmetric versus generally asymmetric matrices reflecting the arrow of time. Such a difference is fundamental and naturally, must be included in any concept of coherence.  Clustering in this setting then reflects the concept of coherence, as a scene that retains its ``appearance", but for now we continue with the idea that maintaining appearance is a sensible idea of coherence.

We proceed to cluster the system summarized by affinity ${\cal W}$ by interpreting the problem as random walks through the weighted {\it directed} graph, $G=(V,E)$ generated by ${\cal W}$ as a weighted adjacency matrix.  Stated equivalently, this is like a directed diffusion problem.  See Sec.~\ref{randomwalkapp} for the comparable discussion in the symmetric case.  So let,
\begin{equation} \label{PfromW}
{\cal P}={\cal D}^{-1}{\cal W},
\end{equation}
where analogously to the symmetric case,
${\cal D}(i,i)=\sum_j {\cal W}_{i,j}, \mbox{ }{\cal D}_{i,j}=0, i\neq j.$  So ${\cal P}$ is a row stochastic matrix, and it represent probabilities of a Markov chain through the directed graph $G$, where,
\begin{equation}\label{probs}
{\cal P}_{i,j}=p(j(t+a)|i(t)),
\end{equation}
and with this in mind, there is an interpretation of this directed graph partition by spectral methods as a naive-Bayes image classifier, by an unknown transfer operator, and  we plan to pursue this perspective in the future; a similar observation that the symmetric diffusion map method relates to a Bayesian update has been made in \cite{CoifmanBayes}. 

{\color{black}Note that ${\cal P}$ is row stochastic implies that {\color{black}it rows sums to one}, or this may be stated  in terms of  the right eigenvector which is the ones vector, ${\cal P}{\mathbf 1}={\mathbf 1}$.  The left eigenvector corresponding to left eigenvalue $1$ represents the steady state row vector of the long term distribution, 
\begin{equation}
u=u {\cal P},
\end{equation}
 which for example if ${\cal P}$ is irreducible, then $u=(u_1,u_2,...,u_{M})$ has all positive entries, $u_j>0$ for all $j$, or say for simplicity $u>0$.}

We may cluster the directed graph by concepts of spectral graph theory for directed graphs, following the {\bf weighted directed graph Laplacian} described by Fan Chung \cite{fanchung}, and a similar computation has been used for transfer operators in \cite{Froyland2,HallerC} and as reviewed \cite{bolltbook}.
The Laplacian of the directed graph G is defined \cite{fanchung},
\begin{equation} \label{LaplacianFormulation}  
{\cal L}=I-\frac{\Pi^{1/2}{\cal P}\Pi^{-1/2}+\Pi^{-1/2}{\cal P}^T \Pi^{1/2}}{2},
\end{equation}
\color{black}
where $\Pi$ is the corresponding diagonal matrix, 
 \begin{equation}
 \Pi=diag(u),
 \end{equation}
 and likewise,
  \begin{equation}
  \Pi^{\pm 1/2}=diag(u^{\pm1/2})=diag((u_1^{\pm1/2},u_2^{\pm1/2},...,u_{M}^{\pm1/{2}})),
  \end{equation}
   which is well defined for either $\pm$ sign branch when $u>0$.
   
\color{black}
   
 See discussion of the symmetric spectral graph theory in, Sec.~\ref{cutSymmetric}-\ref{randomwalkapp}, and the ncut problem solution standard description by Courant-Fischer theory, and how that adapts to 
 this weighted directed graph Laplacian case, in 
Sec.~\ref{directedspectralGraph}.
  
The the first smallest eigenvalue larger than zero, $\lambda_2>0$ such that, 
\begin{equation}
{\cal L}v_2=\lambda_2 v_2,
\end{equation}
 allows a bi-partition, by,
\begin{equation}\label{yv2}
y=\Pi^{-1/2} v_2,
\end{equation}
by sign structure.  As before, analogously to the Ng-Jordan-Weiss symmetric spectral image partition method,  \cite{NgJordanWeiss}, the first $k$ eigenvalues larger than zero, and their eigenvectors, can used to associate a multi-part partition, by assistance of  $k$-means clustering these eigenvectors.

{\color{black}

}


\section{Numerical Results of Motion Segmentation by Time Directed Affinity and Spectral Partition}\label{numericalresults}
{\color{black}
We have  developed in the previous sections a method to find coherent structures from movie data, without trajectories. We described coherent structures to be a set of points that ``hold together" through time (movie frames), but our methodology is designed to infer this concept \textit{absent directly observing particles}.  Now we describe the quality of our results, relative to what we might expect if we had complete knowledge of the system, whether by vector fields or by particle trajectories, although our coherency inference is absent these.  In order to evaluate and compare our numerical results, we review  a performance measure describing the degree of coherency to apply post hoc to our detected coherent structures.

Recall from Eq.\ref{model},  assume we have a dynamical system,
$\dot{{\mathbf x}}={\mathbf F} ({\mathbf x},t),
$
with a vector field, ${\mathbf F}:M\times {\mathbb R} \rightarrow M$. Recall the flow map $x(t) = \Phi(x_0,t)$, that evolves particles from initial position $x_0$. In an autonomous dynamical system, ``hold together" is compatible with the concept of almost invariant sets \cite{Froyland2, bolltbook}, Note that, $A$ is almost invariant if $\Phi(A,0) \approx \Phi(A,T)$, for a time period $T$. 
In \cite{Froyland2} there is described the concept of ``coherent pairs" applicable for non autonomous systems, that allows for pairs of sets to evolve together, and it includes a notion of robustness to break what would otherwise be a truism that any set would be a coherent pair with respect to itself.  Alternatively, we have previously defined \cite{bolltCurv,bolltCurv0} ``shape coherent sets" that describes some sets that may evolve by the flow in a manner that is approximately equal to a much simpler flow, that of a rigid body. That is we minimize, 
\begin{equation} \label{supEq}
{\cal{C}}(A,A) = \sup_{S(A)} \frac{m(S(A) \cap \Phi({A}))}{m(A)}
\end{equation}
over the set of all rigid motions $S(A)$, where $m(\cdot)$ may denote Lebesgue measure.  (More generally, the arguments of ${\cal{C}}(A,B)$ allow for two different sets, as described in \cite{bolltCurv,bolltCurv0}). See Fig.\ref{fig:cohfac}.  This simple theoretical idea is particularly relevant for image analysis because we have a way to measure success empirically by using relatively standard image registration computations. In this manner we will compute numerical values for shape coherence to score each of the following examples.

The shape coherence, Eq.\ref{supEq}, of a set of points $A$ under a flow $\Phi$ through movie frames defines a score, $0 \leq {\cal{C}}(A,A) \leq 1$. The value 1 indicates that $A$ is most strongly  shape coherent through movie frames.  To estimate  ${\cal{C}}(A,A)$ according to contrasting the estimated flow, versus rigid body deformations. We may choose any one of the computed colored coherent sets derived by our directed affinity method. See Fig.~\ref{fig7}-c, and the corresponding same colored regions in Fig.~\ref{fig7}-d (these are our estimations of $\Phi(A)$). Fig.~\ref{fig7} follows our analysis of Jupiter data, discussed in more detail in the next section.
For comparison to rigid body deformations, selecting any one of those  sets from Fig.~\ref{fig7}-c, we optimally register in Fig.~\ref{fig7}-d for estimations of $S(A)$.  Estimated computations of measured intersection of ${\cal{C}}(A,A)$ follow.
Thus, here we have a post-hoc computation  applied to our coherent sets as computed by directed affinity method, whereas in  \cite{bolltCurv,bolltCurv0} relied heavily on the availability of the underlying model such as a vector field to analyze the evolution of boundary curvature.

\begin{figure}[hbtp]
\centering
\includegraphics[trim={20cm 25cm 20cm 25cm},scale=0.05]{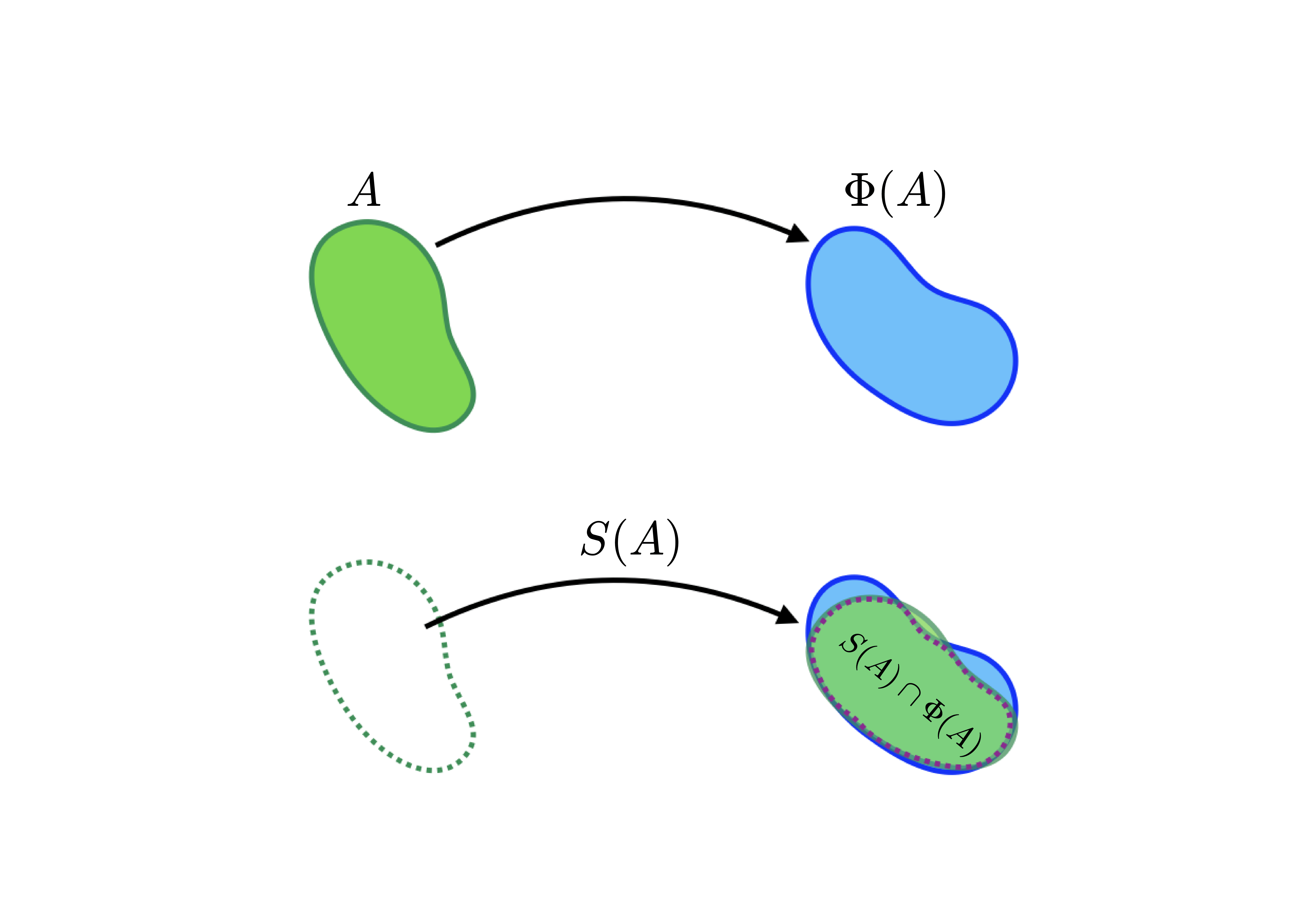}
\caption{{\color{black}For the two sets $A$ and $\Phi(A)$, the intersection of the two sets is found after finding the best geometric transformation that best fit the two sets. We used in our experiments the image registration technique with rigid body registration (translation and rotation).}}
\label{fig:cohfac}
\end{figure}

}
In the next section, we evaluate three example problems, indicating the efficacy of the directed spectral partition method, from our directed affinity from Eq.~\ref{affinityd}-Eq.~\ref{affinity2}.  These will be, the Cassini remotely observed movie of Jupiter, a local lake effect snow event, and a synthetic data set from the double-gyre, in that order.

\subsection{Directed Spectral Partition: Jupiter}
\begin{figure}[ht]
  \centering
\includegraphics[width=1\textwidth]{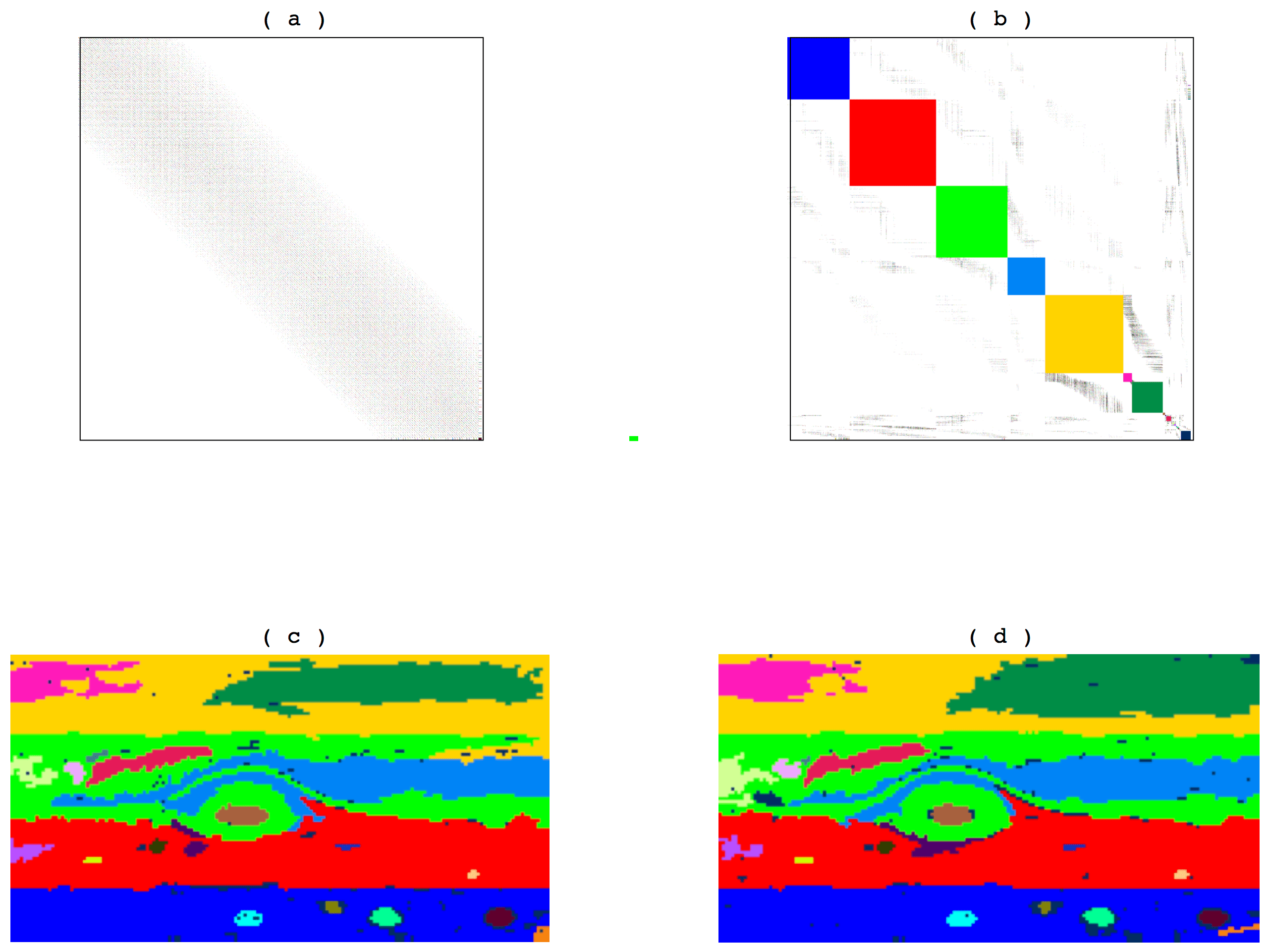} 
  \caption{{\color{black}
Given a small scene surrounding the Great Red Spot, and coarse-grained (for ease of computation and clarity of presentation in this figure), (a) The affinity matrix, ${\cal W}$, Eq.~\ref{affinity2}, (b) Affinity matrix sorted according to spectral partition by methods of Sec.~\ref{directedspectralpartition}, Eq.~\ref{affinity2}-Eq.~\ref{yv2}. (c) Coloring by each block of the sorted affinity matrix partitions the scene according to regions that are found in multiple frames. Eight Frames (1 to 8) were used to create our directed affinity (d) The partitioned scene after $t=T=3$ time. (Frames 4 to 11). }}
  \label{fig7}
\end{figure}
\begin{figure}[htbp]

\centering
\includegraphics[scale=0.5]{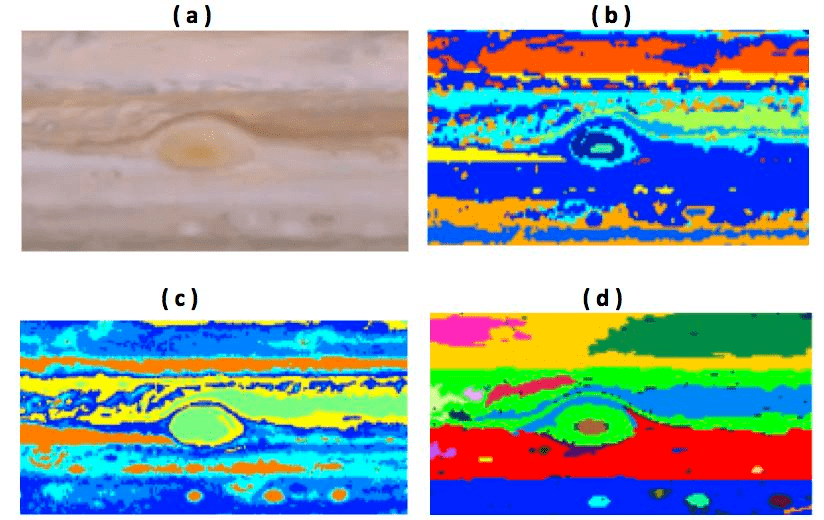}
\caption{(a) A small scene surrounding the Great Red Spot, (b)A k-means clustering based on color only by affinity matrix. (c) Based on spectral partitioning with color alone affinity (d) Based on directed affinity matrix, as in Fig.~\ref{fig7}.}
\label{fig10}
\end{figure}

The results of partitioning using the directed affinity matrix ${\cal W}$ is shown in Fig.~\ref{fig7} from a scene of the GRS, and including the affinity matrix and a permutation that brings it to block structure as indicated by colors matching the colored scene. Fig.~\ref{fig10} again shows a scene of the GRS of Jupiter and its segmentation according to comparing the different methods of k-means to a single scene, a spectral method from a single scene, and finally our directed spectral method.
We see that in our method (d), the regions found by the directed method are most coherent in the sense of showing across time what is clearly visible in a movie, and perhaps difficult to fully appreciate in a static figure here.

Our data set consists of 14 images taken by the narrow-angle camera onboard NASA's Cassini spacecraft. The images span 24 Jupiter rotations between October 31 and November 9, 2000. {\color{black}We refer the reader to} NASA website \cite{NASAImages}, to see how the {\color{black}scene changes} through the movie frames since it is hard to clearly detect the dynamic through still images. In our result, we have chosen a primary number of clusters that maximize the mutual information between movie frames, then, for each cluster, every connected object have been extracted as a separate cluster. {\color{black}We have excluded three frames out of 14 available because they have include significant occlusions appearance of Jupiter's moons within the scene.}

\begin{figure}[htbp]
\centering
\includegraphics[scale=0.45]{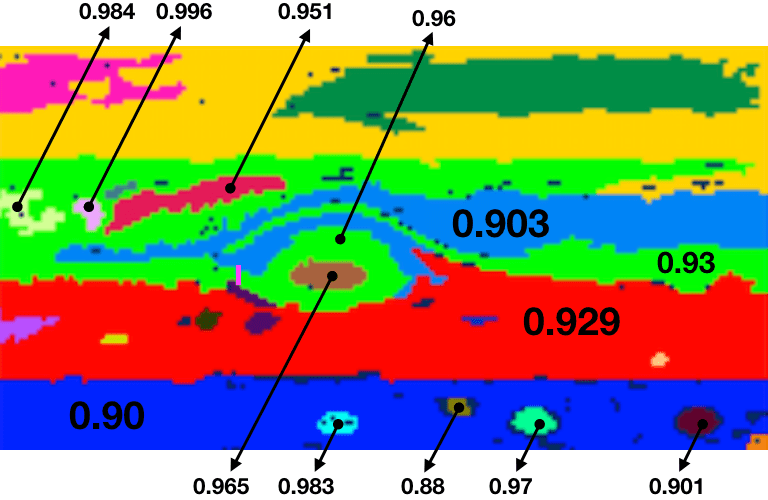}
\caption{{\color{black}Coherence Factor for regions found in Fig~\ref{fig7}. We see that for most regions, the directed affinity methods gives a coherence factor greater than 0.9 found by Eq.~\ref{supEq}, which indicates a high degree of shape coherence. Note that the outer layer of the GRS is connected to the large green region, and to give precise coherence factor for the GRS independently, we made small split shown with small pink line to the left.}}
\label{JupCohFactor}
\end{figure}

\begin{figure}[ht]
  \centering
\includegraphics[width=0.8\textwidth]{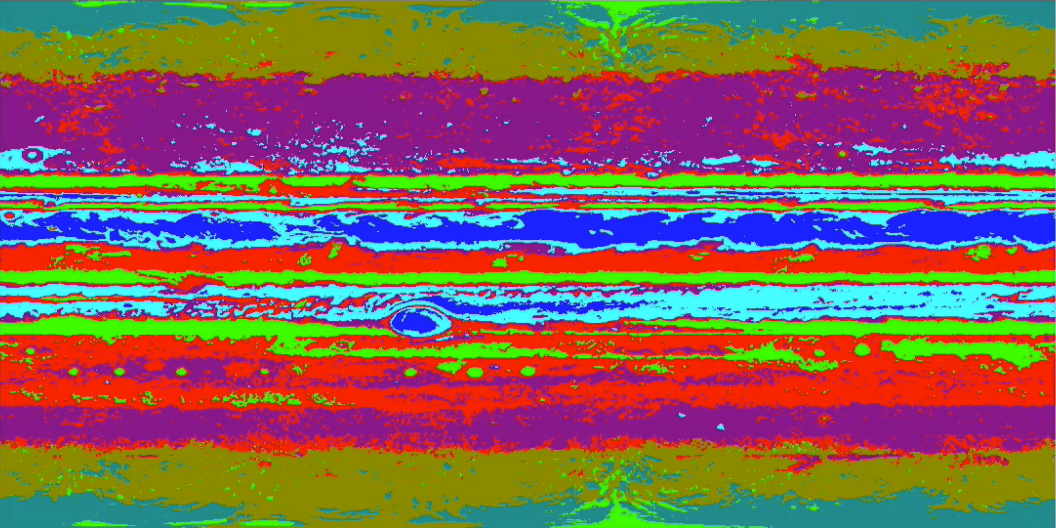} 
  \caption{
  Directed spectral partition of Jupiter of the entire Cassini data set. Compare to Fig.~\ref{fig6}.
 }
  \label{fig6a}
\end{figure}

\begin{figure}[htp]
  \centering
\includegraphics[width=0.6\textwidth]{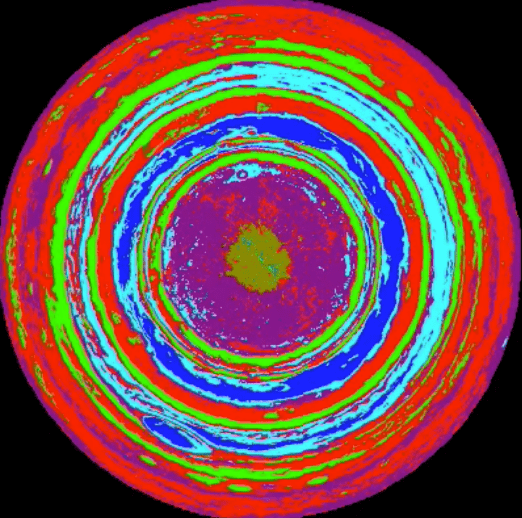} 
  \caption{
  Directed spectral partition of Jupiter as shown on a projection as seen from the northern pole.  Compare to Fig.~\ref{fig6a}.
 }
  \label{fig6}
\end{figure}

{\color{black}
%
Fig.\ref{JupCohFactor} shows the coherence factor by Eq.~\ref{supEq} for the main colored regions in Fig.~\ref{fig7}-c. Note that we see that the directed affinity method detect coherent structures with high accuracy that exceeds 90\%. The Cassini's Jupiter data set is shown in Fig.~\ref{fig6a} as directed spectral partitions for the entire dataset. Most notable are the banded longitudinal structures, the many circular vortex storms, and of course the famous GRS. Also of interest here, the directed spectral partition of the entire data set from Jupiter, as shown from the northern pole. See Fig.~\ref{fig6}. The longitudinal cloud structure can be seen in this global projection to rotate in a manner that reminds us of a twist map\cite{Meiss,TM02,TM03}.  Note the GRS is seen in the 7 o'clock position in this figure.}

\subsection{Directed Spectral Partition: Lake Effect Snow}\label{snowexample}


A lake effect snow is a common scenario during winter months nearby Clarkson University.  It comes from energetic but cold air flowing across a relatively warmer expanses of water, in this case the Great Lake, Lake Ontario.  These local storms are caused by the moist warm air rising into the cold air, that falls as snow nearby as the prevailing airmass sweeps over the colder downwind land.

\begin{figure}[ht]
  \centering
\includegraphics[width=.6\textwidth]{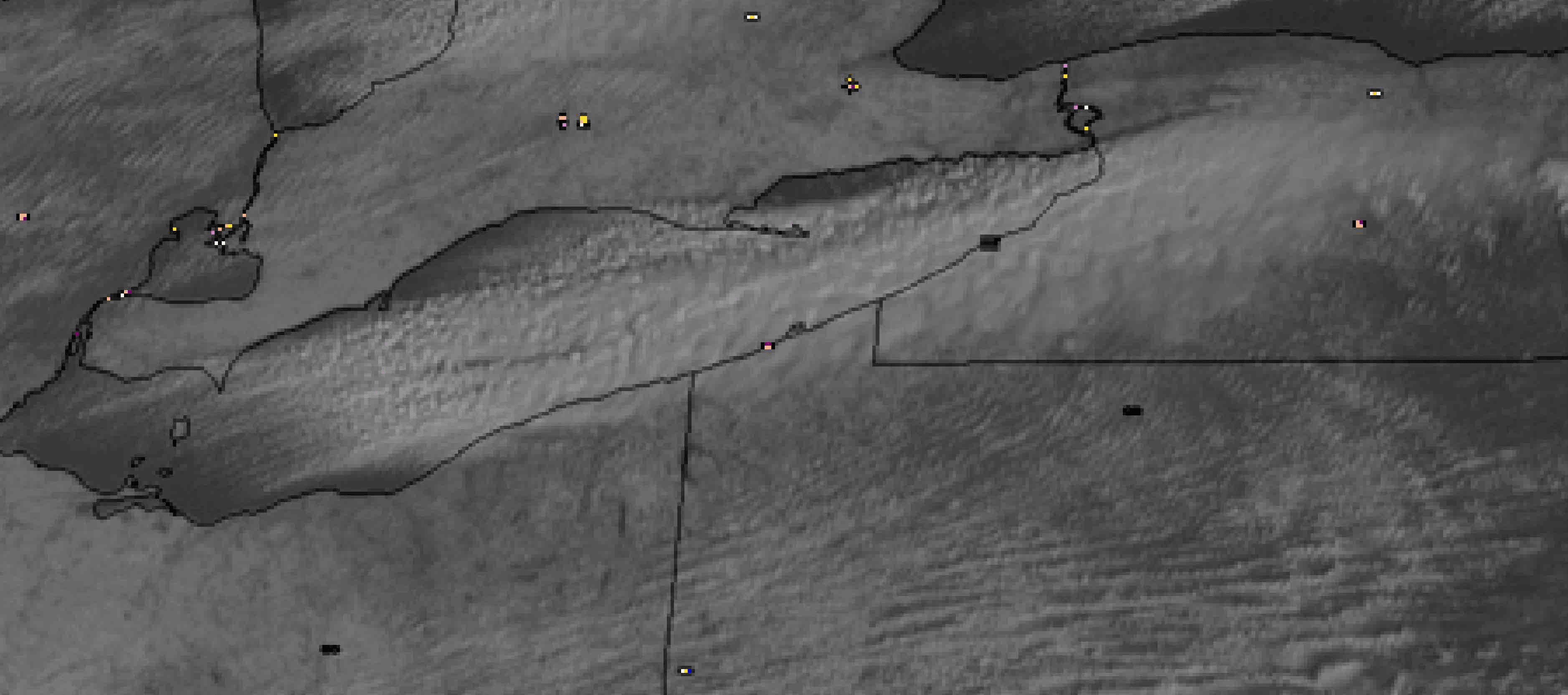}
\includegraphics[width=.6\textwidth]{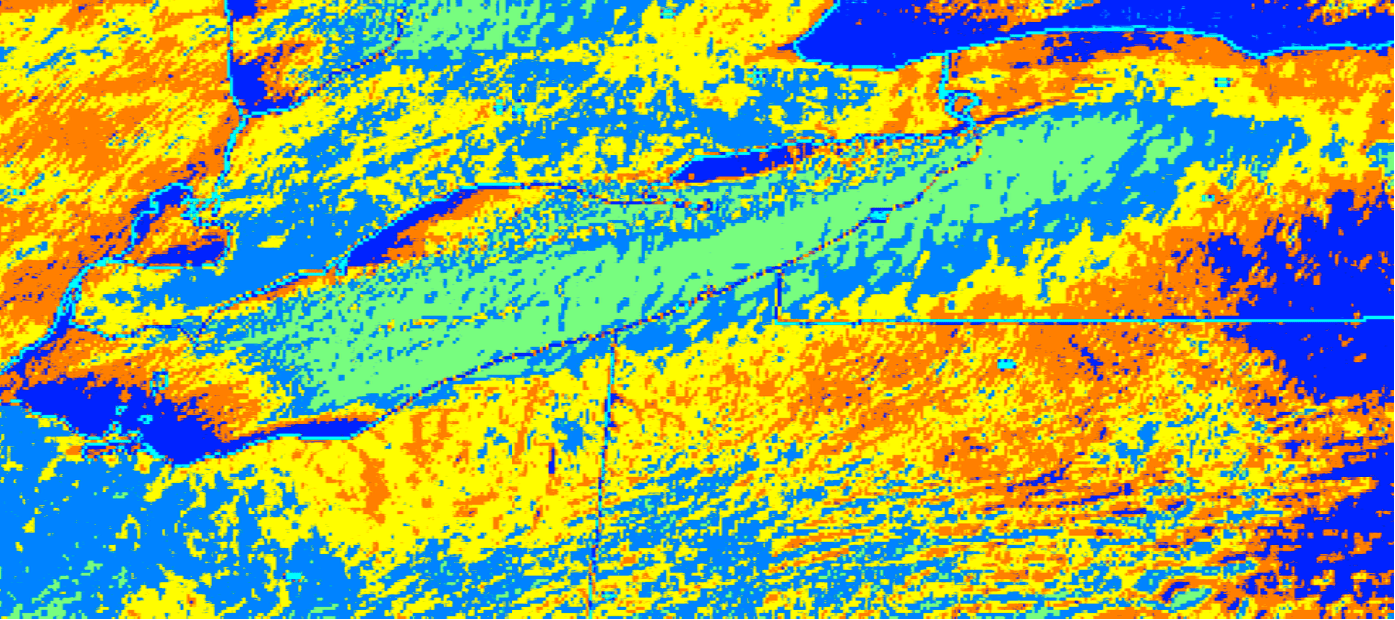}
\includegraphics[width=.6\textwidth]{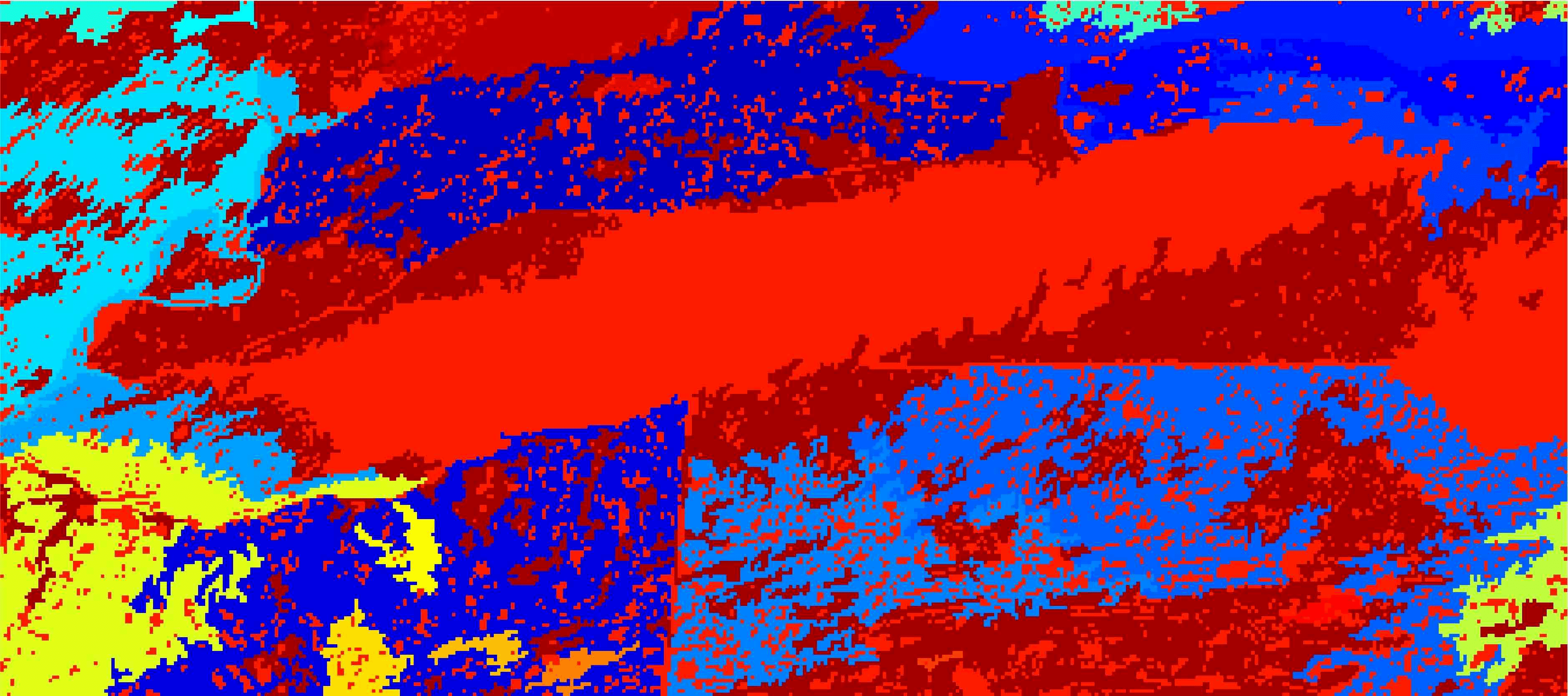}              
  \caption{{\color{black} (Top) Lake effect Satellite Image from SSEC \cite{CCSE,CCSE2}. (Middel) A k-mean clustering based on color by the affinity matrix for the top image. (Bottom) Coherence based on the directed affinity matrix. This image results from 6 time steps - 6 frames - starting from frame number 4 in the raw gif image, and time delay $t=1$. }}
  \label{fig12}
\end{figure}

The hallmark of such an event is a storm that seems to be ``parked", sometimes dropping snow for days in one locale, where even perhaps 50 miles away towns may enjoy sunshine.  Generally, in such regions prone to the events, they happen many times each winter.  They can be persistent and seemingly stationary, lasting for days and dropping massive amounts of snow in a highly localized event, such as for example 8 feet of snow dropped in 10 days over nearby Oswego, NY, in 2007 \cite{nytimes}.  We analyzed one such nearby event for which we had convenient data from 2014. See Fig.~\ref{fig3}-Fig.~\ref{fig12}. In Fig.~\ref{fig12} we see how the static segmentation by affinity matrix (middle) does not reflect any coherent structure while it appears clearly by the directed affinity segmentation (bottom).

The methods herein successfully identify the lake-effect snow storm as a visually apparent coherent structure.  In some ways, storms generally can be described as coherent structures,  expressing energy.  In this case of a lake effect snow, a particular interpretation is interesting.  This is a coherent set that is stationary, even though the underlying flow is advecting, strongly,  from west to east.  

So clearly the coherent structure here is not the outcome of a purely advective process, or even an advective-diffusive process, as assumed in the formulation of most other studies of coherent structures. This one is more akin to the full system which is like an advective-reactive-diffusive process.  

The reactive part is due to the heat bath (literally) associated with the warm lake reacting with the cold advecting air and then later with the even colder land mass downstream. {\color{black} So what we see, and experience, is a derivative of all three aspects of the process.  If we are stationary, as the case of lake effect snow, then the coherent structure is a big deal, and very hard to miss, but not  understood at all in terms of advection alone as normally described in this literature of coherent sets.}

\subsection{Directed Spectral Partition: Double Gyre}

The double gyre system as introduced by Shadden {\it et.~al.} \cite{shadden} has become ubiquitous \cite{bolltbook,haller3} as a benchmark for testing methods for finding or defining coherent structures.
We take the standard version, as a nonautonomous Hamiltonian system, 
\begin{eqnarray} \label{dge}
&& \dot{x}=- \pi A \sin ( \pi f(x,t)) \cos(\pi y) \nonumber \\
&&\dot{y}=\pi A \cos ( \pi f(x,t)) \sin(\pi y) \frac{df}{dx}
\end{eqnarray}
with standard parameters, where $f(x,t)= \epsilon \sin(\omega t) x^2 + (1-2 \epsilon \sin (\omega t)) x$, $\epsilon = 0.1$, $\omega=2 \pi/10$ and $A=0.1$.

\begin{figure}[hbtp] 
\centering
\includegraphics[scale=0.25]{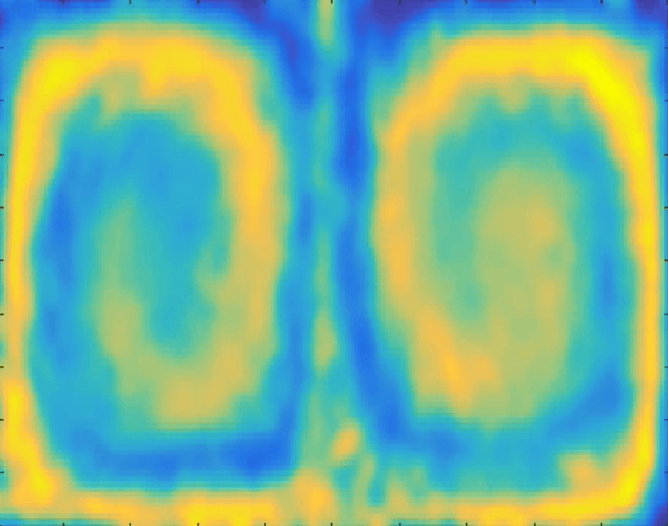}
\includegraphics[scale=0.25]{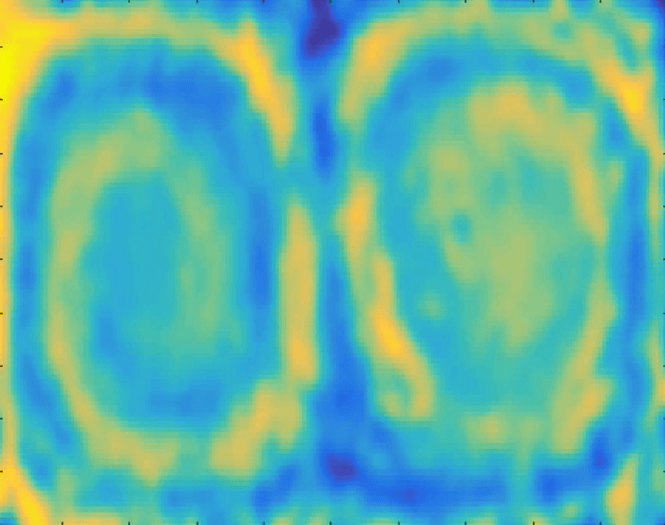}
\caption{{\color{black}Double Gyre snapshots of evolving density profile for ensembles of many initial conditions as a ``movie" used to find coherent sets shown in Fig.~\ref{fig11}. Number of frames is 25 sequential frames, with time delay $t = 2$. }}
\label{dgData}
\end{figure}

\begin{figure}[htbp]
\centering
\includegraphics[scale=0.07]{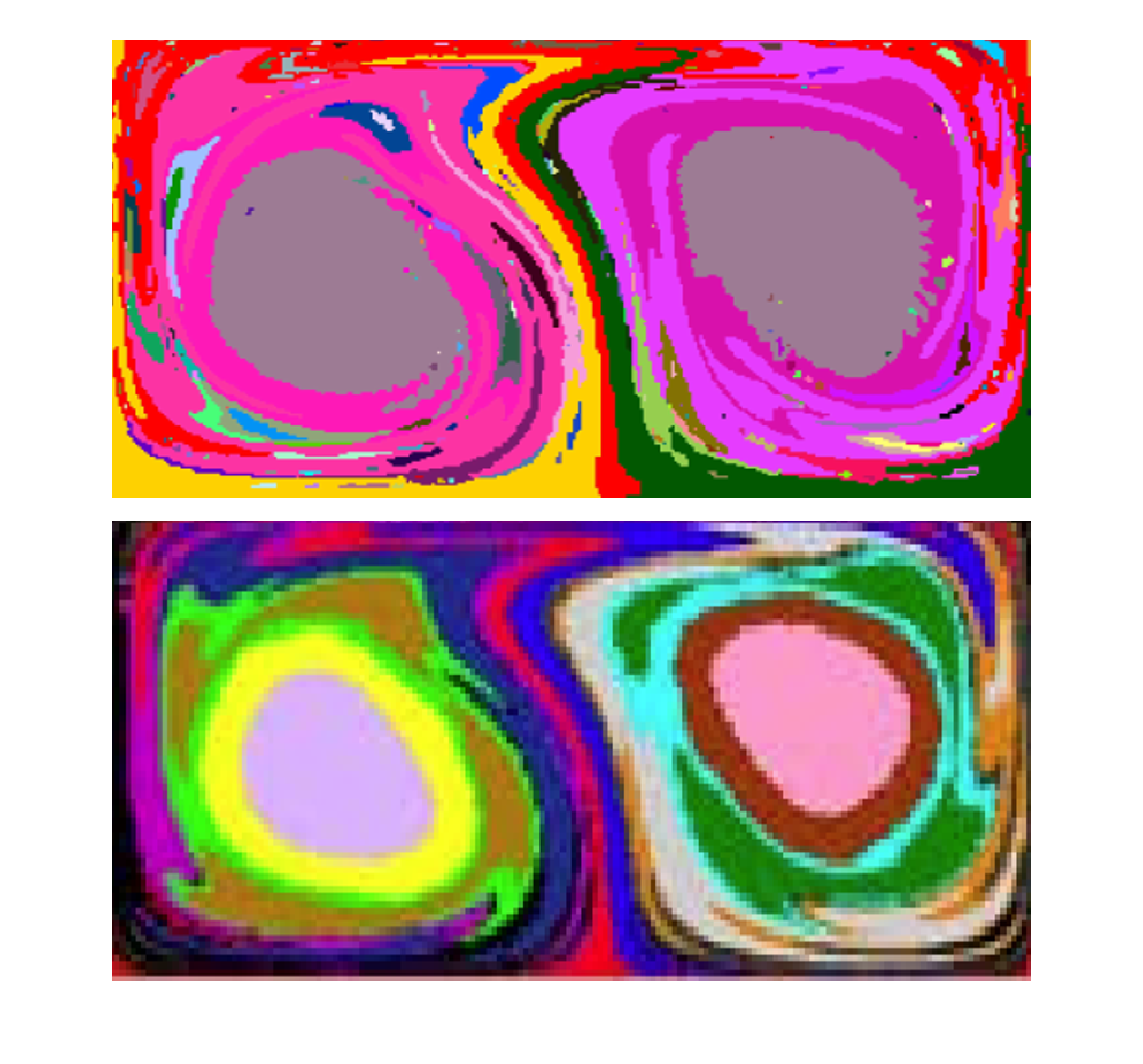}
\caption{{\color{black}Double gyre Eq.~\ref{dge}, (Top) Coherence based on our directed spectral method. (Bottom) Coherent structures found by the true vector field using Frobenius-Perron operator and the Ulam-Galerkin method \cite{TianErik,Froyland6}.}}
\label{fig11}
\end{figure}

\begin{figure}[hbtp]
\centering
\includegraphics[scale=0.05]{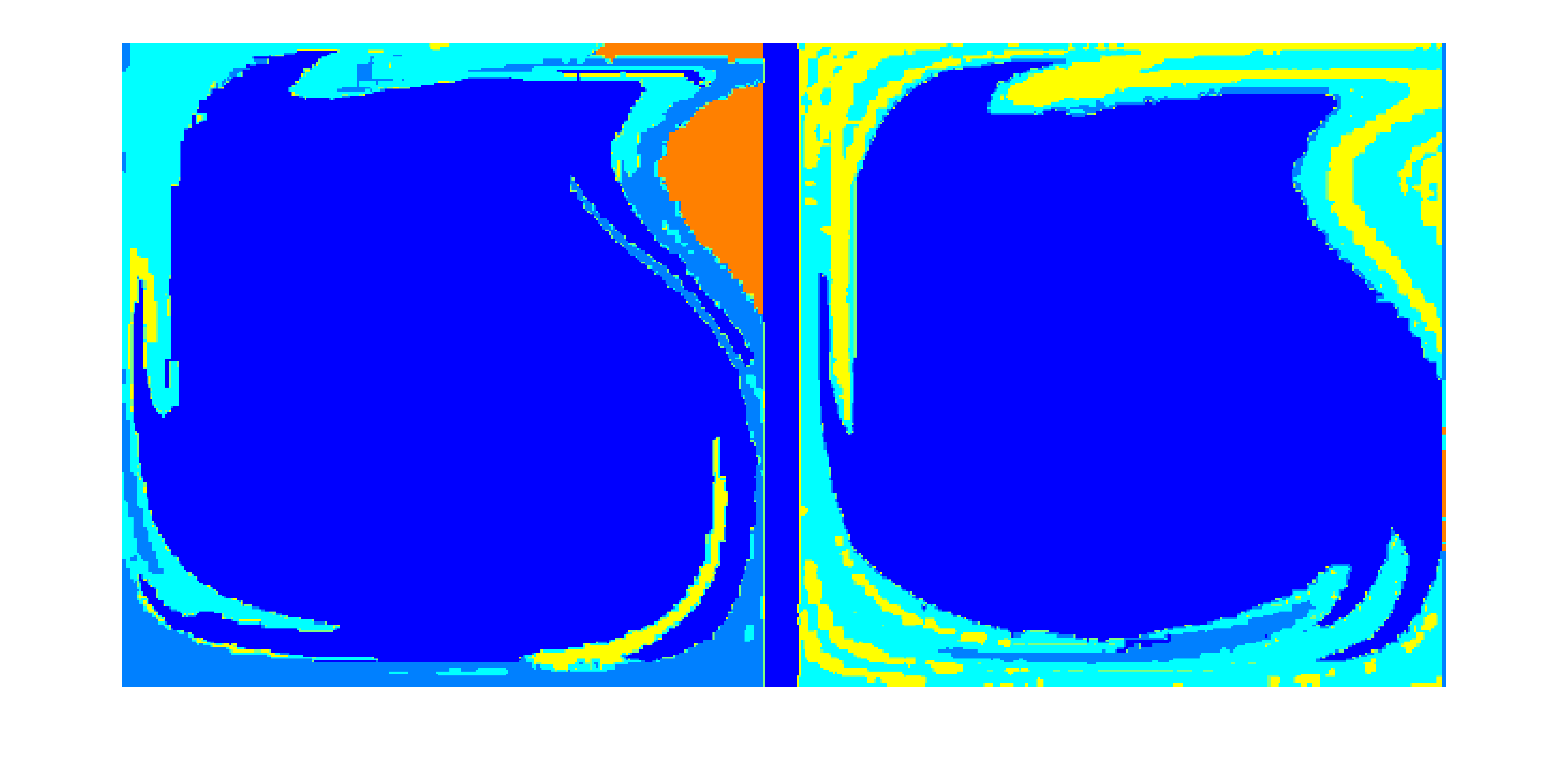} 
\caption{{\color{black}Focusing on the left gyre from Fig.\ref{fig11}: (Left) is computed by our directed affinity spectral method developed here using only the movie as illustrated in Fig.~\ref{fig11}.  (Right) The same region is viewed, and partitioned using methods that relying on knowledge of the model/vector field.  Specifically here have used standard Ulam-Galerkin approximation methods to partition based on construction of the Frobenius-Perron operator, \cite{TianErik,bolltbook}.  Considering our ad-hoc score shape coherence as a coherence factor applied to each of these, we get, $0.955$ for the largest black set shown, and comparable success for the smaller sets.  See also, Fig.~\ref{JupCohFactor} which similarly shows sets scored by this coherence factor.}}
\label{figsLevels}
\end{figure}

\begin{figure}[hbtp]
\centering
\includegraphics[scale=0.05]{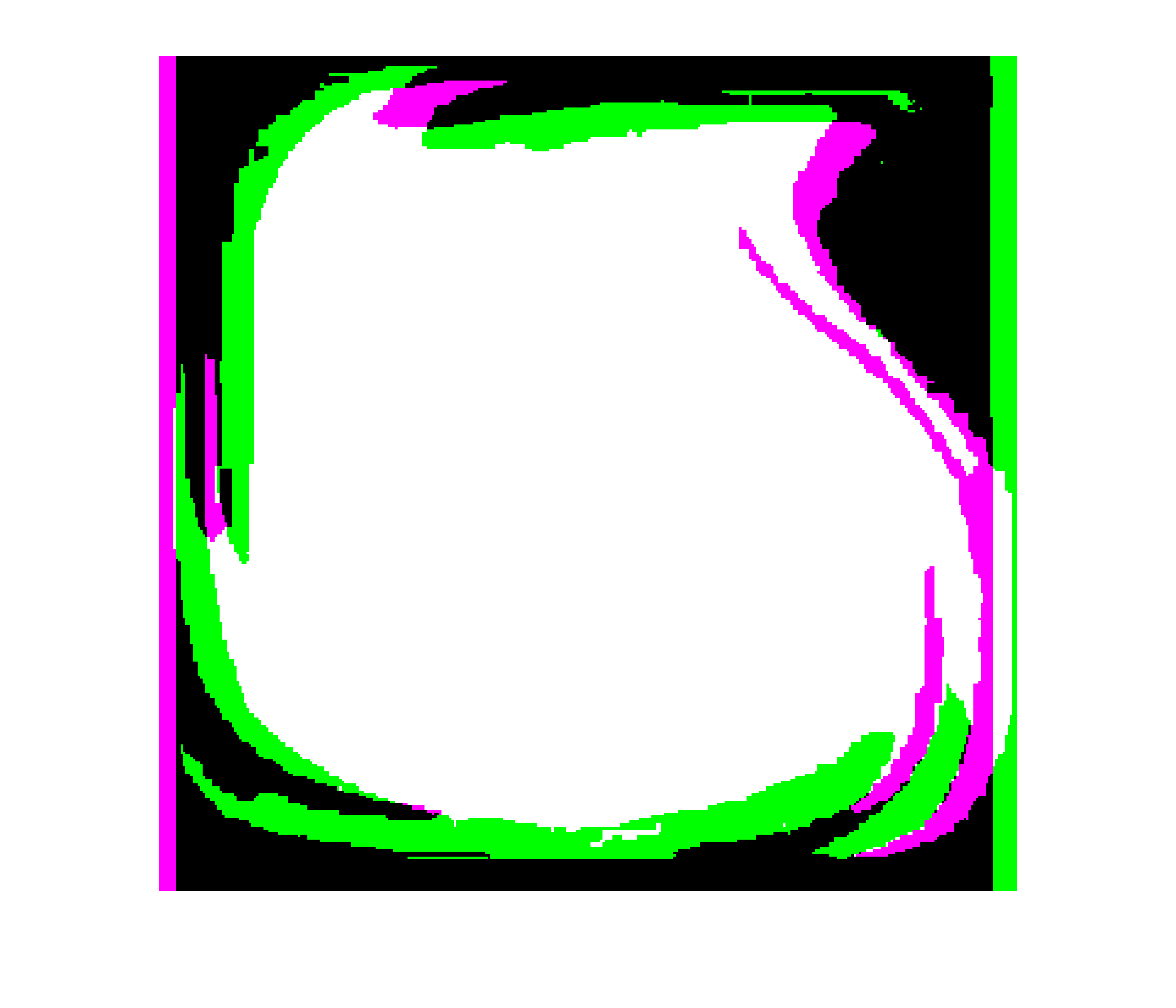}
\includegraphics[scale=0.05]{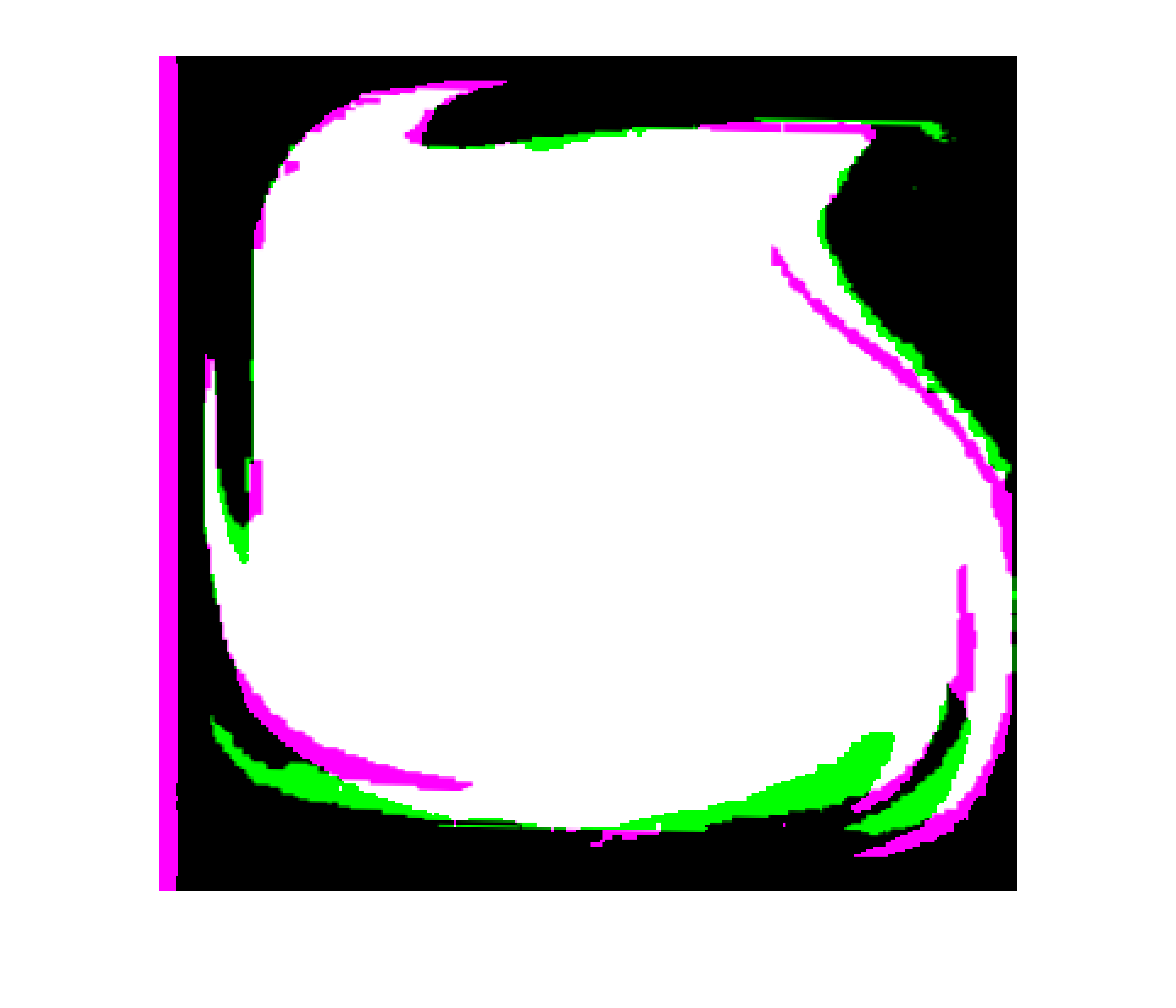} 
\caption{{\color{black}Example of rigid body registration (translation and rotation) for the regions shown in Fig.\ref{figsLevels}. (Left) we see the unregistered images. (Right) Registered images. The geometric transformation of the set $A$ in rigid body registration is found in terms of the translation in $x$ and $y$ directions, and the rotation angle $\theta$. Then, the set $S(A)$ is the set $A$ after applying this geometric transformation.}}
\label{registration}
\end{figure}

{\color{black}Forming a synthetic data set as a movie of evolving density of ensembles of orbits of many initial conditions. We deduce by methods herein  results are clearly similar to many other studies of coherence in the double gyre \cite{Froyland4, Froyland5, HallerB, bolltbook, bolltCurv} by other transfer operators or geometric methods. See Figs.~\ref{dgData},~\ref{fig11}. While the large scale structures are clearly similar to LCS based analysis of coherent sets (particularly the left-right gross-scale partition), but also the appearance of elliptic-island like structures in the middle of the gyres, and the unstable-manifold like structures in the middle between left and right.

Fig.\ref{figsLevels} shows a comparison between the coherent structures found by directed affinity and the coherent structures found by the true vector field using Frobenius-Perron operator and the Ulam-Galerkin method \cite{TianErik,Froyland6}. Here, since we have the true vector field, we can find the coherent structures by direct methods, specifically based on construction of the Ulam-Galerkin approximation of the Frobenius-Perron operator. We compare shown side by side in Fig.~\ref{figsLevels} these ``exact" sets to those from our movie-data-only method developed in this paper.
Our method shows a robustness and accuracy of more than 95\% for different coherent sets.
For image processing optimization of Eq.~(~\ref{supEq}), standard rigid-body registration is used; specifically the command {\it imregister} in Matlab \cite{matlab}, was used as an ad hoc implementation.
Fig.\ref{registration} shows the rigid body registration versus our coherent sets as found by our directed affinity spectra method, from Fig.\ref{figsLevels}-bottom.   }

\section{Conclusion}

We have presented a perspective to infer coherence from remotely sensed ``movie" data.  This is inherently an Eulerian form of a data set since measurements (color) is always associated to a specific location, rather than Lagrangian measurements along orbits.  However, most coherence discussions in the literature are formulated in terms of Lagrangian formulations.  Our methods are inherently spectral in nature, and our details have aspects closely associated with the Meila-Shi spectral image partitioning, but these are also notably similar to the diffusion map methods.  

{\color{black}Since our problem has a defnitive arrow of time as expected for the time-varying aspect of a movie, then the standard symmetric requirement for a spectral method breaks down.} Fortunately the directed graph version of spectral graph theory allows us to handle the weighted directed graphs that we deduce.  We remark that the affinity matrix used in this discussion, ${\cal W}$ from Eq.~\ref{affinity2}, has an interpretation of the symmetric version of an exponential kernel as it appears in the diffusion map literature and this relates to a Bayesian method that yields useful data specification results \cite{CoifmanBayes}. We plan to pursue this interesting feature in the future. 

We have remarked that there are aspects of this question that may associate with a Koopman mode interpretation, but our methods do not resemble the DMD modes analysis since those operators are deduced by a least squares optimization, whereas we hope in the future to interpret  our operators as Bayes estimates that emphasize continuity in space and continuity in time. 

Our observer based perspective in finding coherent sets is based on motion tracking methodology, and it is subjects to motion segmentation limitations when the difference between the sequence of images is too large, where even tracking a solid object become hard and have a high margin or errors.

Our examples have included especially demonstration that the storms and banded structures of Jupiter that are readily apparent by casual inspection.  This simply reflects that our concept of coherence here is more so motion tracking, or motion segmentation than the coherence in literature. By this, we mean that coherence has mostly been associated with advective, or advective diffusive processes.  

However, as we illustrate with the lake effect snow data set, many processes also include a reactive aspect.  As such, this particular data set demonstrates that the measured quantities, such as cloud cover, storm activity, and the like, can remain stationary even while the underlying advection part of the process may be strongly blowing past the process focus.  Our method should be contrasted to standard coherence since by motion tracking we are focusing on measured quantities of interest.  We do not take this to be either a strength or a weakness of the standard methods or our own, but rather we just bring it forward as a point of interest. 

Perhaps there is a connection to the concept of Burning-Invariant-Manifolds (BIM) for reaction diffusion processes \cite{Mitch1,Mitch2,Mitch3}.  While there are other works where spectral methods appear in dynamical systems, including \cite{HallerC,diffusionmap2},
and of course the Koopman methods \cite{rowley}, by DMD and many following papers \cite{appliedkoopmanism}
, and recently connecting Koopman methods and Diffusion maps \cite{Koltai},
as well as the isoperimetric work in \cite{Froyland5}, 
and our own work using community methods from network theory \cite{bolltbook2,BolltModu}, 
the emphasis is on data consisting of tracking Lagrangian particles.
  It is our hope that this work will also serve as a useful further direction to bring spectral methods and clustering methods from data-analytics to dynamical systems concepts of coherence as inferred from real data sets.

\section{Acknowledgments}

This work has been supported by the Office of Naval Research under N00014-15-1-2093, and  the Army Research office under N68164-EG, and the National Geospatial Intelligence Agency.

\appendix
\numberwithin{subsection}{section}

\section{On nCut, The Symmetric Case} \label{cutSymmetric}


Given a graph $G=(E,V)$ generated under the assumption of a $n\times n$ symmetric weight matrix $W$, then a bi-segmentation of the $n$ vertices of $G$ is a bi-partition $v$ and the compliment $v^c=V 
\backslash v$.  Then the standard definition graph theoretic definition of an \textit{ncut} is in terms of volumes of the weighted sets.  Let,
\begin{equation}
vol(A)=\sum_{i\in A} D(i,i), 
\end{equation}
the total weighted degrees from the degree matrix Eq.~(\ref{degreematrix}). This has also been called, $assoc(A,V)$ \cite{Shi-Malik}.
The normalized cut of the partition, labelled $nCut(v)$ of the graph by $v\subset V$ is defined,
\begin{equation}\label{ncutprob}
nCut(v)=(\frac{1}{vol(v)}+\frac{1}{vol(v^c)})\sum_{i\in v,j\in v^c} W_{i,j}.
\end{equation}
where {\color{black} interpreting the sum on the right hand side as a cut,}
\begin{equation}
cut(v)=\sum_{i\in v,j\in v^c} W_{i,j},
\end{equation} denotes the total strength of edges between $v$ and $v^c$.
A ``good" minimal ncut has relatively small weight between the two subsets but strong internal connections.

It can be shown \cite{Shi-Malik} that if {\color{black}$x \in \{-1,1\}^n$ is a characteristic vector} ($n$ dimensional vector with $x_i = 1$ if $x_i$ is in partition $A$ and $x_i = -1$ otherwise, {\color{black} used as an indicator.}) for $v\cup v^c$ then the strong problem $\min_x ncut(x)$ has a relaxation, allowing $y_i \in {\mathbb R}$ rather than $x_i\in\{-1,1\}$,
\begin{equation}\label{ncutprob}
\min_v ncut(v)=\min_{y, y^T D {\mathbf 1}=0} \frac{y^T ({\cal D} - W) y}{y^T D y},
\end{equation}
{\color{black} In other words, thresholding on small values of $y_i$ from a continuum of values allows us to approximately solve the hard threshold problem.}

\begin{proof}
The second part of the equality is a Raleigh quotient that is solved by the eigensystem, 
\begin{equation} \label{eigY}
( D - W)y=\lambda D y,
\end{equation}
as related to,
\begin{equation}
Ly=\lambda y,
\end{equation}
 by Eq.~\ref{eigprob}-Eq.~\ref{eigprobfinish}. Recall that from (3.10) we have:
 \begin{equation}\label{eigsys}
 D^{-1/2}(D-W)D^{-1/2}x = \lambda x
 \end{equation}
 We see that the Laplacian matrix $L=D-W$ is symmetric positive semidefinite, which gives that $D^{-1/2}(D-W)D^{-1/2}$ is also symmetric positive semidefinite and its eigenvectors are pairwise orthogonal, and we can see that $x_{0} = D^{1/2}1$  is eigenvector of Eq.~\ref{eigsys} with $\lambda_{0}=0$ eigenvalue. Then, all other eigenvectors are orthogonal to $x_{0}$. Then:
\begin{equation} \label{constraint}
x_{1}^{T}x_{0} = y_{1}^{T}D1 = 0
\end{equation} 
where $x_1$ is the second smallest eigenvector of Eq.~\ref{eigsys} and $y_1$ is the second smallest eigenvector of Eq.~\ref{eigY}. From the Courant-Fischer theorem\cite{cf} we have,
 \begin{equation}\label{courantFischerFormula}
 \lambda_1 = \min_{x \neq 0 , x \perp x_{0} } \frac{x^{T}Ax}{x^{T}x}
 \end{equation}
with $A=  D^{-1/2}(D-W)D^{-1/2}$ from Eq.~\ref{eigsys}, then we have:
  \begin{equation}\label{courantFischerFormula2}
 \lambda_1 = \min_{x \neq 0 , x \perp x_{0} } \frac{x^{T}D^{-1/2}(D-W)D^{-1/2}x}{x^{T}x}
 \end{equation}
 recall that $x = D^{1/2}y$, so we have:
  \begin{equation}\label{courantFischerFormula3} \nonumber
 \lambda_1 = \min_{y^{T}D1=0 } \frac{y^{T}(D-W)y}{y^{T}Dy}  
 \end{equation} 
 \end{proof}

The relationship of this problem to a random walk is discussed further in Sec.~\ref{randomwalkapp}.

\section{On Random Walks and Affinity}\label{randomwalkapp}

It has been shown \cite{Meila-Shi,Meila1} that partitioning the graph $G=(E,V)$ generated in the case of a symmetric affinity matrix $W$ has a random walk interpretation by developing the reversible stochastic matrix, $P=D^{-1}W.$  This relationship could be interpreted as a major idea behind the diffusion map method  \cite{diffusionmap,diffusionmap2}.

The undirected graph corresponding to the symmetric $W$ of Eq.~\ref{symmetricW} can be interpreted in a diffusion sense {\color{black} as describing probabilities, $P_{i,j}=p(j|i)$ of a random walker } moving to $j$ from $i$. 
For pixels $(x_{k_1},x_{k_2},...,x_{k_r})$ to be grouped as visited by random walkers in that order, according to $W$ of Eq.~\ref{symmetricW}, by $P$ as a Markov chain, we are asking what is $p(x_{k_2},...,x_{k_r}|x_{k_1})$ which equals $\Pi_{i=2}^r p(x_{k_i} | x_{k_{i-1}})$ by independence of jumps in a Markov chain.  {\color{black} Note that we have overloaded the notation in the probability statements to denote the color state at each pixel position.}  In a Markov chain with stochastic matrix $P$, then the eigenvalue problem $Py=\lambda y$ will have a largest eigenvalue $\lambda=1$, and corresponding eigenvector $y={\mathbf 1}$ but the second eigenvector describes strongly connected sets{\color{black};  Meila and Shi \cite{Meila-Shi} showed in the context of random walkers } that minimizing the probability of diffusing between two sets equivalently to the ncut problem Eq.~\ref{ncutprob} of the $W$ matrix, which is useful for connecting concepts of random walks and the spectral graph theory derivative of the graph Laplacian.  The following theorem is supporting evidence relating the two problems.

\begin{proposition}
If $\lambda$ and $y$ are eigenvalue-vector solutions of $Py=\lambda y$ then, $(1-\lambda)$ is an eigenvalue of $(D -W)y=\lambda Dy$ and $y$ is an eigenvector of $(D -W)y=\lambda Dy$.
\end{proposition}

{\color{black}
\begin{proof}
Since $D$ is invertible, the proof follows immediately from,
\begin{eqnarray}
Py&=&(1-\lambda)y. \notag \\
(I-P)y&=&\lambda y  \nonumber \\
D^{-1}(D -W)y&=&D^{-1}\lambda Dy \nonumber \\
(D -W)y &=& \lambda D y \nonumber.
\end{eqnarray}
\end{proof}
}

\section{On Cuts and Directed Spectral Graph Theory }\label{directedspectralGraph}

The problem with using the standard spectral graph theory for partitioning problems, to our scenario of motion tracking, and finding coherent sets, is that our affinity matrices yield not symmetric matrices.  So the discussion in the previous two appendices is not directly applicable.  Fortunately, there is a generalization that can handle our needs for a not symmetric cut problem, stated in Sec.~\ref{directedspectralpartition}.
The  Laplacian matrix of the directed graph from Fan Chung,\cite{fanchung}, in {\color{black}Eq.~(\ref{LaplacianFormulation})} we repeat,
\begin{equation}
{\cal L}=I-\frac{\Pi^{1/2}{\cal P}\Pi^{-1/2}+\Pi^{-1/2}{\cal P}^T \Pi^{1/2}}{2}.
\end{equation}
and from,
\begin{equation}
{\cal L}=I-\Pi^{-1/2}R \Pi^{1/2},
\end{equation}

where {\color{black}$R=\frac{1}{2}({\cal P}+\tilde{P})$, and $\tilde{P}_{i,j}= p_j {\cal P}_{j,i}/p_i$} {\color{black} for all $i,j$, is the time reversed Markov chain so $R$ is the reversiblization, } \cite{bolltbook}.  So analogous to the symmetric ncut problem, Eq.~\ref{ncutprob}, a relaxed, not symmetric ncut problem, can be written, $\min \frac{z^t {\cal L} z}{z^t z}$ subject to, $\sum z_i p_i^{\frac{1}{2}}=0$. The solution of the optimization can be shown by the Courant-Fischer theorem,
{\color{black}\begin{equation}
\lambda_2=\min_{z^t p^{1/2}=0,z\neq 0} \frac{z^t {\cal L} z}{z^t z}=\min_{z^t p=0,z\neq 0}\frac{\sum_{i,j}(y_i-y_j)^2p_i {\cal P}_{i,j}}{\sum_i y_i^2 p_i},
\end{equation}}
attained by the eigenvector $z=v_2$, corresponding to $\lambda_2$ of ${\cal L}$, and where $y=\Pi^{-1/2}v_2$, and $\Pi=diag(p)$ in terms of the dominant eigenvector of $P$.  And thus the spectral partitioning problem is translated to a min-max optimization problem, and for the not symmetric problem, this symmetrization allows the use of the main theoretical tool, the Courant-Fischer theorem that requires a symmetric matrix, as developed in \cite{fanchung}, and reviewed in \cite{bolltbook}.

%

{}
\end{document}